\documentclass[sigconf]{acmart}
\usepackage[utf8]{inputenc}

\usepackage{times}
\usepackage{geometry}
\usepackage{graphicx}
\usepackage{hyperref}
\usepackage{fancyhdr}
\usepackage{titlesec}
\usepackage{tocloft}
\usepackage{setspace}
\usepackage{caption}
\usepackage{subcaption}
\usepackage{listings}
\usepackage{color}
\usepackage{amsmath} 
\usepackage{booktabs}
\usepackage{array}
\usepackage{multirow}
\usepackage{listings}
\usepackage{float}
\usepackage{adjustbox} 
\usepackage{siunitx}
\usepackage{xurl}
\usepackage{paralist, tabularx}
\usepackage{url}
\usepackage{algorithm}
\usepackage{algpseudocode}
\usepackage{booktabs}
\usepackage{tabularx}
\definecolor{lightgray}{gray}{0.9}

\AtBeginDocument{%
  }

\setcopyright{acmlicensed}
\copyrightyear{2018}
\acmYear{2018}
\acmDOI{XXXXXXX.XXXXXXX}
\acmConference[Conference acronym 'XX]{Make sure to enter the correct
  conference title from your rights confirmation email}{June 03--05,
  2018}{Woodstock, NY}
\acmISBN{978-1-4503-XXXX-X/2018/06}

\author{Aditya Bali}
\affiliation{%
  \institution{Ashoka University}
  \city{Rajiv Gandhi Education City}
  \state{Haryana}
  \country{India}
}
\email{aditya.bali_ug25@ashoka.edu.in}

\author{Rupsha}
\affiliation{%
  \institution{Ashoka University}
  \city{Rajiv Gandhi Education City}
  \state{Haryana}
  \country{India}
}
\email{ritwik.ray_ug2023@ashoka.edu.in}

\author{Vidur Kaushik}

\orcid{1234-5678-9012}
\affiliation{%
  \institution{Ashoka University}
  \city{Rajiv Gandhi Education City}
  \state{Haryana}
  \country{India}
}
\email{vidur.kaushik@alumni.ashoka.edu.in}

\author{Anirban Sen}
\affiliation{%
  \institution{Ashoka University}
  \city{Rajiv Gandhi Education City}
  \state{Haryana}
  \country{India}
}
\email{anirban.sen@ashoka.edu.in}

\begin{abstract}
    We present \textit{MediaGraph}, a network-theoretic framework for analyzing reporting preferences in news media through entity co-occurrence networks. Using articles from four Indian news-sources, two mainstream (The Times of India and The Indian Express) and two fringe outlets (dna and firstpost), we construct source-specific co-occurrence networks around the 2020–21 and 2024 Farmers’ Protests. We analyze these networks along three network theoretic axes of centrality, community structure, and co-occurrence link predictability. The link predictability metric is a novel metric proposed that quantifies the consistency of entity associations over time using a GraphSAGE-based model. Our results reveal significant differences in reporting preferences across sources for the same event, and a consistent under-representation of farmer leaders across sources. By shifting the focus from textual signals to relational structures, our approach offers a scalable, label-independent perspective on media analysis and introduces link predictability as a complementary measure of reporting behavior.
\end{abstract}

\begin{document}
\title{MediaGraph: A Network Theoretic Framework to Analyze Reporting Preferences in Indian News Media}
\maketitle


\received{20 February 2007}
\received[revised]{12 March 2009}
\received[accepted]{5 June 2009}


\section{Introduction}
News media has a significant impact on public opinion. The selection, coverage, and framing of socio-political events affects the way citizens think and feel about them. Thus, news reporting preferences of various news-sources have an impact on the democratic processes in a society, including electoral outcomes of a country \cite{budak2016fair,mccombs1972agenda}. Moreover, as posited by the \textit{Theory of Persistent Bias} by Baron et al. \cite{baron06}, the impact of media bias on public opinion often acts as a feedback loop for future reporting around socio-political events, since it is influenced significantly by readers' opinion around certain topics/entities.  

While news media is ideally expected to be objective, reporting on the same event often varies across outlets, resulting in readers being exposed to different aspects, narratives, and entities. Some of these variations arise from legitimate factors such as editorial priorities, audience focus, and the nature or category of the news outlet. However, a substantial body of evidence \cite{haryanto2011media, djankov2003owns} also points to the influence of ownership structures, political affiliations, and other external pressures, which can systematically shape coverage and introduce bias. When individuals rely on only one or two news-sources, such biases can contribute to skewed opinion formation.

In this direction, several previous studies have investigated forms of media bias extant in news media \cite{wang2025media,spinde2021automated}. These studies often categorize media bias into selection bias, coverage bias, and statement bias, based on the attention given to and the style with which a set of aspects and entities are reported by the source. However, most prior studies focus on direct textual analysis of sources using natural language processing (NLP) methods. While NLP-based approaches provide scalable estimates of media bias, they often rely on annotated datasets and proxy signals (e.g., sentiment, lexical choice, or framing cues), which introduce subjectivity in the definition and measurement of bias. Additionally, such models may suffer from limited generalizability across domains, topics, and socio-political contexts. We deliberately avoid using the term \textit{bias} in this study, given its subjective nature and dependence on contextual interpretation across multiple latent dimensions. Accordingly, we employ the notion of \textit{reporting preferences} to describe the patterns observed in our analysis.

We develop a novel network theoretic framework for analysis of reporting preferences of four Indian news-sources. They consist of two mainstream sources, namely The Times of India (TOI) and The Indian Express (IE), and two fringe outlets \footnote{Here, mainstream and fringe outlets are defined based on number of readers of a source, with fringe outlets possessing significantly fewer readers than the mainstream sources.}, namely Daily News and Analysis (dna), and firstpost (FP). We study the reporting preferences of these sources by leveraging the co-occurrence networks of news entities occurring in them. A co-occurrence network is an undirected, weighted graph $G(V,E)$ of nodes $V$ consisting of media entities (e.g., politicians, political organizations, corporations, business-persons, celebrities, and others) that are covered around an event. The edge set $E$ consists of links between these entities based on their co-occurrence in the same article. In other words, if two entities $u$ and $v$ occur in the same news article, an edge/link $(u,v)$ is considered between them. The weight of an edge $w(u,v)$ is the number of times the pair $(u,v)$ co-occurs across news articles for the time period considered. This forms \textit{MediaGraph}, a knowledge graph of entities and the co-occurrence relationships between them for multiple sources around various events.

Our study spans two socio-political events that are related, namely the 2020 \cite{fp1} and 2024 \cite{fp2} Farmers’ Protests, for the four Indian news-sources. The 2020–2021 Indian farmers’ protest was a prolonged national mass movement opposing three agricultural reform laws, while the 2024 farmers’ protest represents a renewed mobilization centered on demands for a legal guarantee of minimum support prices (MSP) and broader agrarian reforms. Both of these protests originated from the states of Punjab and Haryana, and subsequently mobilized large-scale participation across multiple regions, with sustained efforts by farmer unions to march towards the National Capital Region and exert pressure on the central government.

We use Media Cloud, an open repository of news articles from various Indian and International sources to collect articles about the two events, based on various manually selected event specific keywords. Next, we extract named entities from this article corpus, followed by deduplication of these entities using a custom Entity Resolution heuristic. Finally, the co-occurrence network (MediaGraph) for each source and event is formed based on entity co-occurrences in the same article. Leveraging MediaGraph, we compare the news reporting preferences of four Indian digital news-sources, along three primary network theoretic axes:
\begin{itemize}
    \item \textbf{Network Centrality:} We study the difference in reporting of key entities based on their network centralities (eigenvector, betweenness, and Weighted degree centrality). A significant difference in the top ranked entities across sources, for the same event, will indicate a difference in selection and co-occurrence of these entities across sources.
    \item \textbf{Community Analysis:} We analyze the network communities of entities using the Leiden community detection algorithm, and study the most influential entities/leaders within each community in terms of their intra-community PageRank. A significant difference in the leadership of the largest communities across sources, for the same event, will indicate a variation in the co-occurrence patterns among entities based on media coverage.
    \item \textbf{Link Predictability:} Finally, for each news-source, we observe if entity co-occurrence links can be accurately predicted based on a temporal split of the co-occurrence networks. In other words, if for a source, the links between entities for a certain period of the event timeline can be predicted accurately based on the co-occurrence network formed using data from a previous period, we call the news-source highly predictable. A significant variation in link predictability across sources will also indicate a variation in their tendency to cover new or unobvious entity links/co-occurrences. We leverage a graph convolutional network (GCN) called \textit{GraphSAGE} \cite{hamilton18} for this purpose, which learns the properties of network neighborhood of each entity in the training co-occurrence network, and attempts to predict the presence/absence of links between each pair of test entities.
\end{itemize}

Utilizing the three axes of investigation mentioned above, we focus on two primary research questions in this study using MediaGraph: (A) Do Indian news-sources differ significantly in terms of their entity co-occurrence patterns around the same event?, (B) Do the fringe outlets (dna and FP) differ significantly from the mainstream news-sources (TOI and IE) in terms of their entity co-occurrence patterns?

Unlike many previous studies that rely on large amounts of annotated ground truth data, a primary advantage of the proposed approach is its independence on labeled data. Additionally, we propose \textit{link predictability} as a novel metric to quantify news reporting preferences, capturing relational patterns between entities that extend beyond simple measures of selection or standalone coverage.

Our findings indicate that news-sources vary significantly with respect to all of the aforementioned axes of analysis, indicative of their varied reporting preferences for entities, around the same events. While mainstream outlets tend to dominantly cover prominent political actors, fringe sources diversify their reporting through more heterogenous reporting, consisting of non-political and celebrity figures. The reporting preferences also vary between the two protest events aligned to similar social cause and patterns. The only consistent pattern across outlets is the conspicuous under-representation of farmer leaders in the media discourse, despite the movement being primarily driven by them. 

Moreover, our analysis of link predictability reveals distinct reporting patterns in entity co-occurrences across the two protest events. Notably, \textit{firstpost} emerges as the most predictable source, despite being a fringe outlet, followed by the mainstream \textit{The Times of India} across multiple experimental settings. This underscores the value of link predictability as a metric for capturing reporting preferences, as it moves beyond surface-level diversity in individual entity selection to uncover underlying regularities in how entities are systematically connected in media discourse.

Our proposed framework serves as a formative step towards deeper understanding of entity co-occurrence networks in media and their impact on media reporting preferences. As part of future work, we intend to analyze multiple other sources of news, including social media news outlets, regional news-sources, and multimodal content platforms (audio and video based), along the three proposed axes of analysis. We also intend to integrate MediaGraph with our broader goal of creating a news-media monitoring platform for India. Such a platform will effectively aid the Indian news media in self-regulation, thereby ensuring its movement towards a more equitable reporting of issues and entities around socio-political events.

\section{Related Work}
Previous studies on analysis of media bias and reporting preferences primarily span two directions of research: (A) Content Analysis and (B) Network Analysis around media data. Here, we review a subset of these studies and highlight how our approach extends and complements their findings.
\subsection{Content Analysis around Media Bias}
A significant number of previous studies have analyzed different forms of biases in mass media data, using traditional content analysis methods. These methods leverage automated analysis of coverage and sentiment around topics \cite{daoudi2026media,sen2022analysis}, automated linguistic similarity across topical \cite{hosseinmardi2025unpacking}, word, and document embeddings \cite{liu2026war}, and qualitative content \cite{budak16} and discourse analysis \cite{shahid2025mapping,belghoul2025media,agustian2025analyzing}. While many of these studies focus on an aggregate analysis of variance in media discourse around the same event, some target automated analysis of ideological bias around named entities mentioned in media articles \cite{9381344,hamborg2021newsmtsc}, including studies \cite{yogevmeasuring} that use multiple supervised and unsupervised NLP-based methods, to analyze coverage, selection, and framing bias around named entities in mass media.

Most of these studies, while bringing out crucial insights, have a dependency on traditional NLP-based methods like topic modeling, sentiment and stance classification. These methods often rely on surface-level lexical signals, making them underperform under paraphrasing and stylistic variations, and are typically constrained by predefined labels or lexicons that fail to generalize across domains, topics, and cultural contexts. Some others purely relying on manual qualitative analysis, inherently are limited by subjectivity and lack of scalability.

To move beyond the limitations of purely text-based NLP and purely qualitative methods, we develop a mixed-methods approach of leveraging entity co-occurrence networks from media articles, and analyzing them using centrality measures, community detection, and GCN-based link prediction. In this approach, bias is reflected not just in isolated word/phrase choices, but in how entities are associated and grouped together across news-sources. While the quantitative analysis grounds our findings in structural patterns within media co-occurrence networks, the qualitative component enables us to interpret these patterns, contextualize them within broader narratives, and derive meaningful insights about the nature of bias across sources.

There also exist research attempts at LLM based content analysis for bias detection. However, most of these studies propone the superiority of smaller, domain fine-tuned classical models over much larger LLMs \cite{horych2025promises}. Additionally, several studies have focused on the inherent biases in these models, which get amplified over time and progressive synthetic data ingestion in the training pipeline \cite{wang2025bias}. 

\subsection{Network Analysis around Media Bias}
There have been fewer studies around network theoretical analysis of news media bias, compared to bias detection using content analysis based methods. Some of these studies focus on word co-occurrence networks and analysis of topical biases using them \cite{https://doi.org/10.1111/soc4.12779}. Others analyze polarization in news media user communities by studying user networks formed around comments and impressions towards digital news-sources \cite{lee2025network}, and synthetic signed social networks \cite{LOW2022126722}. Prior work also models news ecosystems on social media using user-content and interaction networks, showing that polarization and media bias emerge from network structure, community formation, and influence dynamics, rather than text alone \cite{cicchini2022news,flamino2023political}. Unlike these studies, we analyze entity co-occurrence networks in Indian news media, and attempt to analyze these networks along the axes of entity centrality, communities, and link predictability. 

We draw our motivation for this study from \cite{traag2016structure} where the authors show that entity co-occurrence networks in media are highly non-random and self-reinforcing, i.e., prominent individuals not only appear more often in media, but also disproportionately and repeatedly co-occur with other prominent actors, leading to dense, elite-centric clusters. They further demonstrate that media systematically reinforces existing entity associations, thereby shaping a structured and unequal visibility landscape in news coverage. We extend this work in two ways. First, our analysis studies the applicability of its findings in the Indian context, for two mainstream and two fringe news outlets. Second, we add an additional axis of link prediction in entity co-occurrence networks, which serves as a proxy for predictability in entity coverage in news reporting.

\section{Data Collection}
We collected data for the four news-sources for the two protests (2020-21 and 2024) from Media Cloud based on manually decided keyphrases as shown in table \ref{tab:keyphrases}. These keyphrases were finalized after multiple rounds of data collection, manual analysis of the articles, and keyphrase augmentation.
\begin{table}[h]
    \centering
    \begin{tabular}{|p{3cm}|p{5cm}|}
    \hline
    \textbf{Protest Timeline} & \textbf{Keyphrases}\\
    \hline
        2020-21 & "farmers protest" OR "farm laws" "delhi chalo" OR "kisan andolan" OR "singhu border" OR "tikri border" OR "agricultural reform" \\
        \hline
        2024 & "farmers protest" OR "delhi chalo" OR "msp guarantee" OR "minimum support price" OR "shambhu border" OR "samyukta kisan morcha" OR "farmers march"\\
        \hline
    \end{tabular}
    \caption{Keyphrases used for data collection for the 2020 and 2024 protests}
    \label{tab:keyphrases}
\end{table}

We utilized two main APIs for data collection:
the \textit{Media Cloud API} and the \textit{Newspaper API}. The Media Cloud API enables retrieval of article content using customized parameters such as keywords/phrases and dates. By querying its archival collections, it provided us with a list of URLs containing the keyphrases. Using this initial dataset, we then utilized the Newspaper API to parse each URL, and collect the title and the body text for each article. We then cleaned the data and removed certain non-essential fields (e.g., Language, Category, etc.),
retaining only the essential columns in a structured format: Publish Date, Title, Media URL, and the first 100 words of each article’s parsed text. The first 100 words were retained following the inverted pyramid style of journalism, which states that most of the essential information in a news article is generally contained in the beginning (usually the first paragraph)  \cite{po2003news}.

Finally, we obtained 7271, 3454, 538, 878 articles for TOI, IE, dna, and FP, respectively for the 2020-21 protests. For 2024, we obtained 301, 516, 52, and 39 articles for TOI, IE, dna, and FP, respectively. This data was considered for further analysis.

\section{Methodology}
We describe in this section the methods used in this study. We start with discussing the process of entity extraction and resolution, followed by the structural analysis performed on MediaGraph. Finally, we discuss the details of link prediction on the 2020-21 protest graph for the news-sources.
\subsection{Entity Extraction}
We extract named entities from the news articles using  a standard named entity recognition algorithm \cite{honnibal2020spacy}, more specifically spaCy's \verb|en_core_web_trf| model. Since our further analysis deals with entities of type \textit{Person}, we also identify the entity types (POL for politician, DIR for businessperson, BUR for bureaucrats, and ORG for organizations) using an existing dataset \cite{CHEN2025103471} of Indian politicians, directors, administrative service officers (IAS), and companies (listed in Bombay Stock Exchange). The matching and merging of the entities collected from news articles with the existing entity dataset required us to perform efficient searching and matching of the large dataset of entities.

For this purpose, we leveraged \textit{Elasticsearch} \cite{elasticsearch2018elasticsearch}, a popular search and analytics engine built on \textit{Apache Lucene}. Elasticsearch offers robust full-text search capabilities, allowing us to efficiently clean, normalise, and search the data. We set up an index for the unresolved (not deduplicated) entities, storing three primary fields: (A) The entity’s name, (B) Possible aliases of the entity, (C) The entity’s type (POL for Politicians, DIR for Directors, ORG for organizations/corporations) where available. The unresolved index is initialized with all of the entities collected from news articles, and the entities present in the existing dataset (with their types). The \textit{Aliases} field initially contains the name of the entity as obtained. We next proceed to matching and merging of these entities based on the similarity in their names.

\subsection{Entity Resolution}
Entity Resolution (ER) or deduplication is the process of identifying whether multiple records refer to the same entity, and merging them to ensure disambiguation of the data under consideration. 
Any standard ER process consists of two steps:
\begin{itemize}
    \item \textbf{Matching:} This step involves calculating pairwise record similarities by matching record attributes (e.g., name, age, etc. while matching records belonging to people).
    \item \textbf{Merging:} This step involves merging records with above threshold similarity together to form clusters of resolved records. In our case, merging updates the \textit{Aliases} field of the entities, by adding more name variations to it.
\end{itemize}
The matching step, to avoid incorrect mergers, must also ensure that the matching criteria (or similarity measure) is sufficiently strong. The result of the ER process is a group of record clusters, where records belonging to the same cluster are assumed to refer to the same entity. For instance, if the original unresolved collection contains the entity records (in the order [Name, Aliases, Type]):\\\\
$[\textit{John Doe, [John Doe], POL}]$ \\ $[\textit{John D, [John D], \_ }]$,\\\\ 
Upon completion of the ER process, the updated record becomes [\textit{John Doe, [John Doe, John D], POL}]. As we can see, the \textit{Aliases} field now contains both of the aliases, and the type field is updated using the entity type from the first record. 

Our custom ER heuristic deals with resolving the names of popular news entities, which are mentioned frequently in national news sources (e.g., politicians, companies, directors of companies, and bureaucrats). 
ER around media data is a daunting task, owing to the sheer scale of articles, and also because the same entities are mentioned in multiple ways in different articles and outlets. The following examples instantiate a few of the difficulties that are to be handled by the ER algorithm: 
\begin{itemize}
    \item The records \textit{Bhupinder Singh} and \textit{Bhupendra Singh} refer to distinct entities (with name spellings slightly differing), and should not be merged in the same record cluster. If these are merged, a \textit{false hit} error is produced, indicating incorrect merging. 
    \item The records \textit{Lalu Prasad} and \textit{Lalu Prasad Yadav}, however, refer to the same entity (a highly prominent politician named "Lalu Prasad Yadav") and should be merged. If such records remain unmerged after the ER process runs, a \textit{false miss} error is produced, indicating an incorrect miss of a merger. 
\end{itemize}

This algorithm employs a deterministic, rule-based strategy that first preprocesses the entity names, by converting them to lower case and removing possessives, punctuation, and honorifics. 
Next, it proceeds as follows: Let $n_1$ and $n_2$ be two preprocessed names. The algorithm applies the following six rule based filters to match the two names:

\begin{algorithm}
\caption{Entity Resolution Algorithm}
\label{alg:ER}
\begin{algorithmic}[1]
\Require Pre-processed names list $\mathcal{E}$
\Ensure Clusters $\mathcal{C}$

\State $\mathcal{U} \gets \mathcal{E}$
\State $\mathcal{C} \gets \emptyset$

\While{$\mathcal{U} \neq \emptyset$}
    \State Remove any name $p$ from $\mathcal{U}$
    \State $S \gets \{p\}$
    
    \ForAll{$e \in \mathcal{U}$}
        \If{Match($p$, $e$)}
            \State Add $e$ to $S$; remove $e$ from $\mathcal{U}$
        \EndIf
    \EndFor
    
    \State Add cluster $S$ to $\mathcal{C}$
\EndWhile

\State \Return $\mathcal{C}$
\end{algorithmic}
\end{algorithm}

\begin{itemize}
    \item \textbf{Abbreviation subset matching}: This filter ensures that the sorted set of initials for one name must be a subset of the other ($abbrv(n_1) \subseteq abbrv(n_2)$ or $abbrv(n_2) \subseteq abbrv(n_1)$). For example, the names $narendra$ $tomar$ and $narendra$ $singh$ $tomar$ are considered for further matching since the first set of initials $\{n,t\}$ is a subset of the second set $\{n,s,t\}$. However, $narendra$ $modi$ and $narendra$ $tomar$ are not considered for further resolution. This step prevents erroneous matchings, ensuring significant reduction the ER time complexity. 
    \item \textbf{Minimum two-token requirement}: 
    This filter ensures that the names contain at least two words to be considered for further resolution. Thus, we exclude single word names/surnames like $singh$ and $thakur$ from the ER process.
    \item \textbf{Fuzzy first-name matching}: This filter ensures that the first names of the two names in the pair must have a Levenstein similarity of above 0.75. Usually, the first names are mentioned before surnames in Indian media. This step filters out name pairs which do not match significantly in the first names.
    
    \item \textbf{Fuzzy surname matching}: This filter ensures that the surnames of the two names in the pair have a Levenshtein similarity of above 0.8.

    \item \textbf{Phonetic agreement via DoubleMetaphone}: This filter ensures that the phonetic similarity (primary codes) of the two names $\{n1,n2\}$ are equal.
    We use the \textit{DoubleMetaphone} function from the \textit{Metaphone} library to encode both strings phonetically, checking if they sound similar.
\end{itemize}

As a special case, tokens with `ji' suffixes were stemmed as it is a gender-neutral honorific used frequently used for public figures in India (for instance, \textit{Narendra Tomar Ji} was converted to \textit{Narendra Tomar}). 
The thresholds indicated in the above rules were experimentally determined through iterative testing of the algorithm.


Algorithm \ref{alg:ER} shows the custom ER algorithm used. The algorithm iterates over each entity name $p$, and performs a pairwise match with every other entity in the unresolved set $\mathcal{U}$. Names that exhibit a similarity score above a certain threshold (with the original entity name $p$, in this case experimentally determined at 0.85), are clustered/merged together with $p$ (updated as aliases) and stored in the set $\mathcal{C}$. Finally, the set $\mathcal{C}$ is returned as the resolved set of clusters.

To uniquely identify each resolved name cluster, we associate a \textit{canonical name} as a representative for the cluster, which is usually the longest name variant in the resolved cluster (e.g., the alias set [john doe, j. doe, jon doe, jon d.] has the canonical name "john doe" as the cluster representative). 


\subsubsection{Qualitative Performance Analysis} 
To assess the performance of the ER heuristic, we performed a manual analysis for \textit{false hit} errors. Two independent annotators first checked the entity clusters for each news-source for false hits (names incorrectly merged), by labeling each cluster with a 1 (indicating at least one \textit{false hit} within the cluster) or 0 (no false hits). False hits were detected after a manual check of the entity names from the news articles and other external sources. This exercise revealed a high inter-annotator agreement of 98\%. The disagreements were then resolved through one round of deliberation.

To calculate the percentage of false hits, we first found the percentage of incorrect entities in each resolved cluster (ratio of the count of invalid entities and total number of entities within a cluster), and then averaged them across the clusters. This method yielded an accuracy of 97.11\% for TOI, 94.08\% for IE, 98.26\% for dna, and 95.16\% for FP with an error rate $<=6\%$. This indicates a high performance achieved by the rule-based entity resolution pipeline based on false hits. 

To calculate the percentage of false misses, we used the formula:
\begin{equation}
    FMP_N = \frac{FN_N}{FN_N+TP_N}
\end{equation}
where the subscript $N$ denotes a news-source, FMP is the fraction of false misses, FN is the number of entity pairs incorrectly missed (not resolved), and TP is the number of entity pairs correctly resolved.

The task of detecting all possible pairs of entities that should be resolved was significantly difficult, owing to the time complexity of $O(n^2)$ for $n$ entities. To mitigate the complexity of the task, annotators partitioned the entities into alphabetical bins (based on their alphabetically sorted abbreviations), restricting the search for relevant entity pairs to these specific subsets. The total number of false miss errors were minimal ($<6\%$ for all news-sources). 

We went ahead with the ER algorithm discussed in this section. As part of future work, we are planning to improve the algorithm through machine learning based techniques and multi-pass methods like R-Swoosh \cite{benjelloun2009swoosh}, to make the pipeline further efficient.

\subsection{MediaGraph Preparation}
To analyze the entity co-occurrence network for various news-sources, we constructed a weighted undirected graph $G = (V, E)$ (termed as the \textit{MediaGraph}) from the resolved entity co-occurrences (i.e., for an entity name, its canonical/cluster representative name as updated by the ER process was considered), removing self-loops to avoid trivial predictions. Here, $V$ represents the set of PERSON entities  and $E$ represents the set of edges signifying entity co-occurrence in at least one news-article. The edges are timestamped (storing the first and last co-occurrence dates based on the publication dates of the articles), and weighted to capture the total number of co-occurrences in the time period considered. We prepare two separate knowledge graphs for the first and second Farmers' Protests, individually for each news-source analyzed.  

\subsection{Structural Analysis of MediaGraph}
We analyze MediaGraph with respect to various structural properties at the node level, which we define in this section.
\subsubsection{Weighted degree Centrality}



The weighted degree centrality calculates the relative weight of an edge corresponding to a node. It is represented as:\\

$ C_{weighted}(i) = \frac{\sum_j{w_{ij}}}{|E_{ij}|}$\\\\
where $C_{weighted}(i)$ represents the weighted degree centrality of node $i$, $w_{ij}$ is the weight of the edge $(i,j)$, and $E_{ij}$ represents the edges connecting node $i$ to its neighboring nodes $j$.

The weighted degree centrality is then min-max normalized with respect to all nodes to map it within the [0,1] range. The weighted degree quantifies the average frequency with which an entity co-occurs with other entities, serving as a proxy for its overall prominence or omnipresence within political discourse. While we do consider this metric in our further experiments, we do not report the centrality plots since they are very similar to the eigenvector centrality findings.
\subsubsection{betweenness Centrality}
This metric measures how often a node lies on shortest paths between other nodes in the knowledge graph. Thus, entities with high betweenness often may act as \textit{bridges} connecting different network communities formed around the political narrative. 
Mathematically, this is denoted as

$$ C_{betweenness}(i) = \sum_{j<k} \frac{|p_{jk}(i)|}{|p_{jk}|} $$

where $p_{jk}(i)$ denotes the set of shortest paths from node $j$ to $k$ through node $i$, while $p_{jk}$ denotes the set of shortest paths between all pairs of nodes.

\subsubsection{eigenvector Centrality}
This metric highlights nodes that are influential not simply because they have numerous connections, but because they are linked to other highly central or influential nodes in the network. A high eigenvector centrality in MediaGraph thus reveals an entity embedded at the core of the discourse, playing a pivotal role in sustaining and amplifying the its structural backbone. Mathematically, this is denoted as:

$$C_{eigen}(i) = \frac{1}{\lambda} \sum_{j} A_{ij} x_j$$

Where $A_{ij}$ denotes the adjacency matrix entry indicating whether entity $i$ is connected to entity $j$, and $x_{j}$ is the centrality score of node $j$.
The term $\lambda$ is the largest eigenvalue of the adjacency matrix, ensuring the scores are scaled consistently.

\subsubsection{Community Detection}
To get an idea of the communities of entities for each news-source, we use the \textit{Leiden Community Detection Algorithm} \cite{traag2019louvain}. This technique guarantees well-connected, high-quality clusters through iterative refinement and local moving of nodes. This method allows us to track how groups of related actors emerge and evolve over time, offering sharper insights into shifts in narrative focus and the interplay of influential figures throughout the protest. Each entity node gets a community ID as a feature, which is used further in the downstream task of link prediction.

\subsection{Link Prediction on MediaGraph}
We measure the predictability of entity co-occurrence links for each news-source in MediaGraph, and perform a comparative analysis along this axis across sources. In other words, given an undirected entity co-occurrence graph created for a certain span of time, we want to observe if we could predict co-occurrence links between entities for a future time-span. This measure of link predictability loosely connects to the notion of \textit{selection bias}, defined as a statistical bias that occurs when the process of selecting a sample from a population is not random and results in a sample that is systematically unrepresentative of the population. Consequently, a higher link predictability for a source compared to others, during the same time-span for an event, indicates its propensity to select co-occurrence links between entities embedded within structurally similar neighborhoods, and vice versa.

It is important to note that a news-source having a higher link predictability according to our approach does not indicate its lack of novelty in reporting. Nor do we intend to comment on the quality of reporting through this measure. In fact, there can exist newsworthy events which `demand' the reporting of similar set of entities over time, in accordance with appropriate journalistic practices, leading to a higher link predictability. Here, we compare four newspapers reporting on the same event, in turn capturing their differential reporting of entity linkages during the same time span. The approach merely acts as a formative step towards assessing the entity reporting priorities of these sources. Future work can aim to look deeper into the various characteristics that contribute towards novelty in news reporting. \\

We capture \textit{link prediction} in media co-occurrence networks through two approaches -- supervised and unsupervised -- involving \textit{GraphSAGE}, a graph convolutional neural network. GraphSAGE employs \textit{SAGEConv} layers that aggregate neighborhood information through message passing, creating node representations that capture multi-hop relationships. The architecture offers key advantages over classical GCNs through its inductive learning framework, enabling models to generalize to unseen nodes and new graph structures. To enable a systematic evaluation of link predictability, we designed our \textit{Basic Setting} around link prediction using both supervised and unsupervised methods. Additionally, under an \textit{Extended Setting} we also evaluated our methods by augmenting structural network features (like centrality metrics and community identifiers) with Node2Vec node embeddings capturing the network neighborhood of nodes. Finally, for each experimental setting, we employed both incremental/rolling and static training-test splits. These methods are discussed in detail next.
\subsubsection{Node2Vec Input Embeddings}
We first train \textit{Node2Vec} \cite{grover16} using the entity co-occurrence network graphs, in an unsupervised manner using random-walk-based neighborhood prediction, learning general structural representations independent of the downstream task. We use a relatively large walk length of 100, and a high number of walks per node (300), to ensure rich coverage of the graph, along with a context window of 15 to capture higher-order co-occurrence patterns and broader structural relationships in the entity network\footnote{These parameters were experimentally determined, based on the final performance in the GraphSAGE-based link prediction task}. These embeddings are fed as input to the GraphSAGE model, both for supervised and unsupervised link prediction.

While Node2Vec and GraphSAGE both capture node embeddings based on structural properties of a graph, by initializing it with Node2Vec embeddings, we supply GraphSAGE with a rich structural prior that encodes global context, allowing the model to focus on refining these representations for the downstream task rather than learning them from scratch. This was particularly beneficial in our relatively sparse co-occurrence graphs, where purely random or one-hot vector initialization led to slower convergence and suboptimal representations. Thus, our approach can be seen as a hybrid representation learning strategy\footnote{We separately tried using only Node2Vec (without GraphSage) in the link prediction task. However, the results were sub-optimal, exhibiting the superiority of the hybrid approach.}, combining unsupervised Node2Vec structural encoding with GraphSAGE-based link prediction.

\subsubsection{GraphSAGE-based Supervised Link Prediction} \label{sec:sup}

\begin{figure*}[t]
    \centering
    \includegraphics[width=1\linewidth]{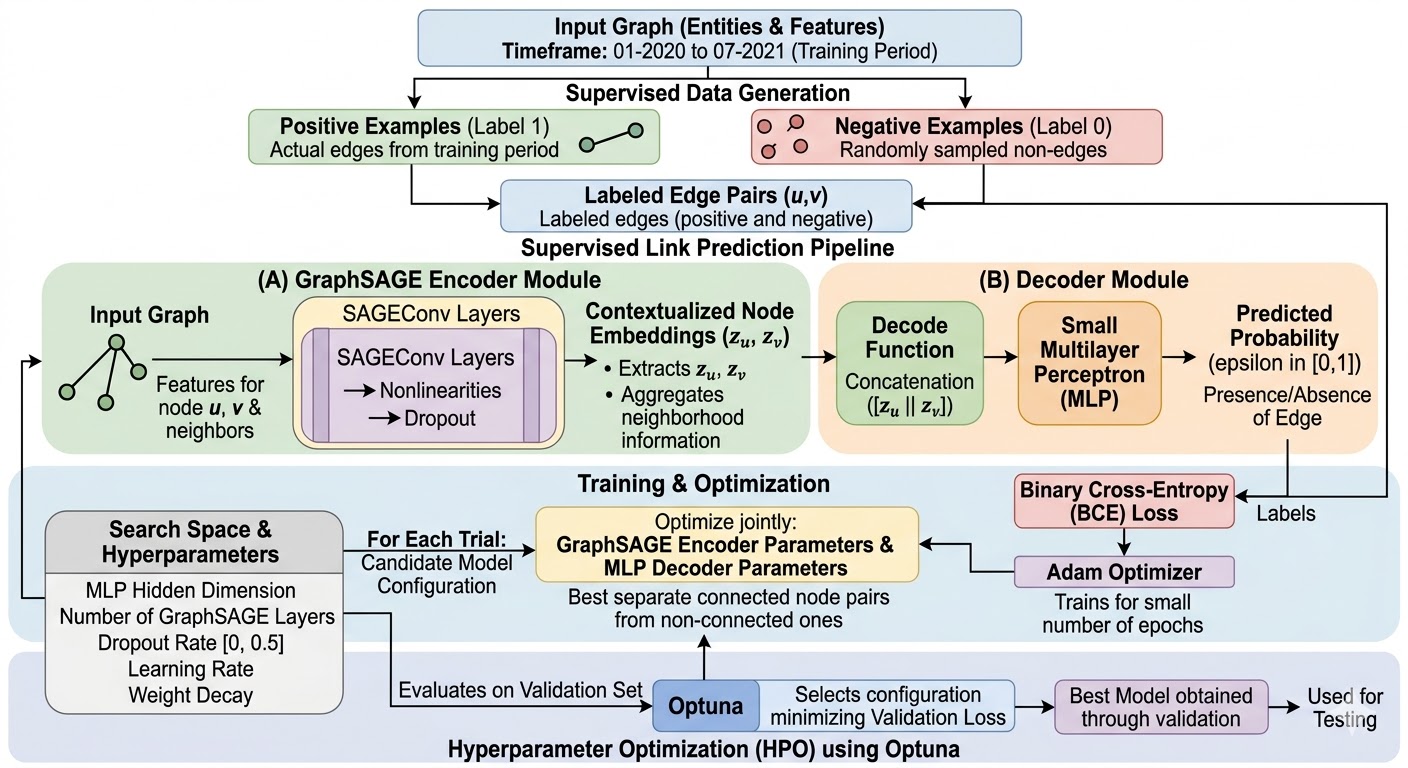}
    \caption{The supervised link prediction approach that trains the encoder and decoder modules together on the downstream link prediction task. In the unsupervised case, the encoder and decoder are decoupled; the encoder is trained on an unsupervised loss function, while the decoder is separately trained on link prediction.}
    \label{fig:supgcn}
\end{figure*}

The supervised approach (figure \ref{fig:supgcn}) under the basic setting uses labeled edges between entities as the ground truth. A binary label of 1 or 0 indicates whether there exists an edge between an entity pair or not. The positive examples (labeled 1) are actual edges from the training period and negative examples are randomly sampled pairs of vertices with no link between them.

The pipeline consists of two stages: (A) A GraphSAGE encoder module, and (B) A decoder module. The encoder applies a stack of SAGEConv layers with nonlinearities and dropout, producing contextualized embeddings that aggregate information from each node’s neighborhood. It outputs a pair of node embeddings $(z_u,z_v)$ for each input pair $(u,v)$. For supervision, the decode function concatenates the two embeddings (along with some other structural features that we explain in table \ref{tab:inputs}) output by the encoder, and feeds the resulting vector $z_u||z_v \in R^{2d}$ through a small feed forward network that outputs a probability ($\epsilon[0,1]$) indicating the presence/absence of an edge. 
where $d$ is the dimension of each feature vector. During training, the model receives labeled edges (positive and negative), optimizes a binary cross-entropy loss over these pairs, and learns embeddings and parameters that best separate connected node pairs from non-connected ones. Therefore, in this setup, the node embeddings are task-specific: they are learnt jointly and directly for the end goal of link prediction.
We tune the supervised model using \textit{Optuna}, which searches over key hyperparameters such as the hidden dimension for the MLP, number of GraphSAGE layers, dropout rate, learning rate, and weight decay. For each trial, a model is instantiated and trained for a small number of epochs using the \textit{Adam} optimization using binary cross-entropy loss. The model outputs link probabilities through the feed forward network decoder, with dropout (within range [0,0.5]) applied after each GraphSAGE layer to reduce overfitting. After training, the candidate model is evaluated on the validation set, and Optuna selects the configuration that minimizes the validation loss. Finally, the best model (learning rate of $3 \times 10^{-3}$, final dropout of 0.3) obtained through the validation is used for testing. The approach is visualized in figure \ref{fig:supgcn}.

\begin{table}[t]
\small
\centering
\begin{tabularx}{\linewidth}{p{3cm}|X}
\hline
\textbf{Input} & \textbf{Description} \\
\hline

Node2Vec Embeddings &
64-dimensional Node2Vec embeddings (input in both basic and extended link prediction) \\

Node Type &
Categorical attributes such as entity type (PERSON/POL/ORG) encoded as a scalar value \\

Community ID &
Community IDs obtained from graph clustering algorithms, encoded as a scalar value \\

Centralities &
Normalized numerical features such as eigenvector, betweenness, and weighted degree centralities, encoded as scalar values \\
\hline
\end{tabularx}
\caption{Summary of node embeddings and structural features fed to the GraphSAGE link prediction model as input.}
\label{tab:inputs}
\end{table}
\subsubsection{GraphSAGE-based Unsupervised Link Prediction} \label{sec:unsup}
The unsupervised approach under the basic setting uses similar input features across experiments as the supervised approach, with some differences in the training methodology. Contrary to the supervised approach, here, the node embeddings are first learnt independently of the downstream task of link prediction, by optimizing a random-walk–based unsupervised objective (encoder); only after this pretraining step are the resulting embeddings fed into a feed forward link classifier (decoder) for supervised link prediction, thereby decoupling representation learning from the end task. We describe the encoder and decoder modules below.

\textbf{Encoder:} For the GraphSAGE encoder, a pair of entity node embeddings with a rich feature set (Node2Vec embeddings and other structural features) is fed as input, similar to the supervised approach. The model then learns embeddings for the input nodes using an unsupervised loss function defined based on the GraphSAGE paper by Hamilton et al. \cite{hamilton18}. 

This model learns node representations by maximizing the similarity of observed (positive) edges, while pushing apart randomly sampled negative pairs. Negative sampling follows the standard unsupervised GraphSAGE formulation: for every positive pair $(u,v)$, we draw $Q$ random nodes to construct hard negative examples and penalize their similarity to the anchor node $u$. 

For negative sampling (node pairs that are originally not directly connected in the training data), we use a 50-50 mix of random pairs and two hop negatives. Two-hop negatives provide us node pairs where the two nodes are connected via two hops, but are not directly connected. 
These samples provide harder examples that force the model to learn subtle distinctions beyond simple connectivity. 



Finally, for each observed training edge $(u,v)$ and each negative edge $(u,v_n)$, we compute the dot-product similarity of the corresponding node embeddings, and optimize a logistic loss that encourages high scores for positive edges as shown below: 
\begin{align*}
\mathcal{L} = & -\frac{1}{|E_{\text{train}}|} \sum_{(u,v) \in E_{\text{train}}} w(u,v) \log \sigma(\mathbf{z}_u^T \mathbf{z}_v) \\
& - \frac{Q}{|E_{\text{train}}|} \sum_{(u,v) \in E_{\text{train}}} \mathbb{E}_{v_n \sim U(V)} \log \sigma(-\mathbf{z}_u^T \mathbf{z}_{v_n})
\end{align*}
where $\sigma$ denotes the logistic sigmoid, $E_{train}$ denotes the edge set in the training data. As seen from the loss function above, the positive-loss term is averaged across all observed edges, while the negative-loss term is averaged across all randomly sampled negative pairs, thereby yielding empirical Monte-Carlo approximations of the expectations appearing in the original unsupervised GraphSAGE objective.

Training for the unsupervised encoder proceeds with Adam optimization and an adaptive learning-rate scheduler (\textit{ReduceLROnPlateau}), together with a patience-based early-stopping criterion to prevent overfitting. After convergence, the encoder is run in evaluation mode to obtain the final node embeddings ($z_u$ and $z_v$) used in downstream link prediction experiments.

\textbf{Decoder:} We construct node-pair features similar to the supervised approach, by concatenating the learnt embeddings ($z_u$ and $z_v$) from the encoder. 

A two-layer feed forward network with architecture $R^{2d} \rightarrow R^{h} \rightarrow R^2$ serves as the final link predictor, where the hidden dimension $h \in \{64, 96, 128, 192\}$ is optimized using Optuna. We employ dropout (0.3) and early stopping to prevent overfitting. The output layer comprises two units representing a binary classification task, where the model predicts the probability of the presence (label 1) or absence (label 0) of a link between a given pair of nodes.

\subsubsection{Extended Setting: Employing Structural Features}
We implement both the supervised and unsupervised approaches using a basic setting (one that uses only concatenation of node embeddings as input in the encoder as discussed in sections \ref{sec:sup} and \ref{sec:unsup}) and an extended setting (where network structural features described in table \ref{tab:inputs} are incrementally added to the input, alongside the node embeddings). In the extended setting, we concatenate the community ID, three centrality values, and node type scalars along with the node embeddings, resulting in an input embedding of dimension $R^{2d+5}$. Thus, we get four experimental set-ups that we summarize in table \ref{tab:experiments}. These network structural features are expected to provide global and meso-scale signals about a node's position and importance in the graph that are not always captured by local message-passing alone. This helps the model distinguish nodes that may look similar locally, but occupy very different structural roles, thereby improving the quality and discriminability of the learnt embeddings.

\begin{table}[t]
\centering
\begin{tabularx}{\columnwidth}{p{3cm} p{5cm}}
\hline
\textbf{Experiment} & \textbf{Description} \\
\hline

1A: Basic Supervised Link Prediction &
Supervised link prediction using only node embeddings as input, establishing the baseline model performance. \\
\hline
1B: Basic Unsupervised Link Prediction &
Unsupervised link prediction using negative sampling, with only node embeddings as input. \\
\hline
2A: Supervised Link Prediction using Structural Features &
Supervised link prediction using node embeddings and structural features (see Table~\ref{tab:inputs}). \\
\hline
2B: Unsupervised Link Prediction using Structural Features &
The unsupervised analogue to Experiment~2A, where each structural feature group is appended to the node embeddings. \\

\hline
\end{tabularx}
\caption{Summary of the experimental variants evaluated in the link prediction pipeline.}
\label{tab:experiments}
\end{table}

\subsubsection{Temporal Evaluation and Edge Weight Threshold}
The configurations (experiments 1A , 1B, 2A, and 2B) were systematically evaluated under two temporal settings: (A)\textit{Incremental/Rolling Training Approach}, in which the model is trained on all data available up to a selected cutoff month (August 2021) and subsequently evaluated on each following month, approximating real-world, time-ordered deployment (rolling window), and (B) \textit{One-time Training Approach}, which involved training and validating the model on the training and validation data (up to July 2021), and evaluating it collectively across the links existing for the remainder of the year (August to December, 2021). These settings are detailed in table \ref{tab:training}. 
Additionally, we applied edge-weight thresholds\footnote{Minimum number of co-occurrences that qualify as an edge.} to capture only the statistically significant co-occurrences in the training data (number of co-occurrences >= 2). In the following section, we report our findings using a combination of these experimental settings.

\begin{table}[h!]
\centering
\begin{tabular}{|p{1.5cm}|p{2cm}|p{2cm}|p{2cm}|}
\hline
\textbf{Config.} & \textbf{Training Period} & \textbf{Validation Period} & \textbf{Testing Period} \\
\hline
Incremental Training &
Rolling (Jan 2020 to $t$ where $t \in$ [May 2021 to October 2021]) &
Rolling (one month post training timestamp) &
Rolling (one month post validation timestamp) \\
\hline
One-Time Training &
Jan 2020-May 2021 &
Jun 2021-Jul 2021 &
All links formed/existing between Aug 2021 to Dec 2021 \\
\hline
\end{tabular}
\caption{Experimental regimes: incremental vs. one-time training, with training, validation, and testing periods.}
\label{tab:training}
\end{table}

\section{Results}
In this section, we present our findings around the analysis of data for Farmers' Protests (2020-2021 and 2024), derived from MediaGraph. 
The co-occurrence network statistics for the four news-sources considered are presented in Table ~\ref{tab:basicstats} shows the number of nodes and edges in the entity co-occurrence networks for the sources over two periods: a 24 month period from January 2020 to December 2021, and a 6 month period from January to June 2024 (only PERSON entities considered). We can see that the mainstream news-source TOI forms the largest network, followed by FP, IE, and dna. Additionally, the 2020-21 networks are significantly larger than the 2024 network, owing to the smaller scale of the second protest and the data collection having been completed in 2024. 

We also observe a small value of network density of the networks for all four news-sources for both of the protest events, indicating significant sparsity. However, the fringe outlets (dna and FP) show a significantly higher density than the mainstream sources (TOI and IE) for both protests. This indicates that fringe outlets tend to concentrate their coverage around a smaller, more tightly interconnected set of entities, leading to denser co-occurrence structures. In contrast, mainstream sources cover a broader and more diverse set of entities, resulting in larger but sparser networks. We now report our findings along the three axes of network centrality, community analysis, and link prediction.

We acknowledge that the results derived from the entity co-occurrence networks of dna and FP for the 2024 protests are preliminary, as these networks are extremely small in size. Such sparsity may arise from limitations in data collection (e.g., via Media Cloud) or from the specific time window during which the data was collected. Consequently, any observed patterns are unlikely to be robust or generalizable.


\begin{table}[h]
\centering
\small
\begin{tabular}{|c|c|c|c|c|}
\hline
\textbf{Year} & \textbf{Source} & \textbf{\#Nodes} & \textbf{\#Edges} & \textbf{Density} \\
\hline

\multirow{4}{*}{2020--21} 
& TOI       & 6197 & 14014 & 0.000730 \\
& IE        & 1533 & 3373  & 0.002872 \\
& dna       & 865  & 5091  & 0.013624 \\
& FirstPost & 2822 & 59340 & 0.014908 \\

\hline

\multirow{4}{*}{2024} 
& TOI       & 252 & 466 & 0.014735 \\
& IE        & 299 & 483 & 0.010842 \\
& dna       & 33  & 38  & 0.071970 \\
& FirstPost & 18  & 14  & 0.091503 \\

\hline
\end{tabular}
\caption{Summary statistics of entity co-occurrence networks across news sources, for PERSON entities}
\label{tab:basicstats}
\end{table}

\subsection{eigenvector Centrality}
The eigenvector Centrality analysis in figure \ref{fig:ev_comparison} reveals interesting insights into the structural influence of entities, i.e., those connected to other highly influential nodes within the media narratives of the 2020-21 and 2024 Farmers' Protests.

We find that both of the mainstream outlets (TOI and IE) maintain a fundamentally political elite-centric network structure across both protest events. In both outlets, \textit{Narendra Modi} consistently ranks as the entity with the highest eigenvector centrality score, reflecting his omnipresence in the discourse's structural backbone. This trend is similar to that followed by the key opposition figure \textit{Rahul Gandhi}, who is one of the top ranked entities in both protests, for the mainstream outlets.

In 2020-21, dna also shows presence of political figures among entities with high centralities (like \textit{Narendra Modi} and agriculture minister \textit{Narendra Singh Tomar}). However, we also see presence of popular non-politician entities like \textit{Greta Thunberg} (international activist) and \textit{Rihanna} (international celebrity) in the ranking. This structure presents a mix of political, activist, and celebrity presence in the discourse, which is different from that of mainstream outlets. Preliminary findings from the limited data in 2024 show that dna's top influential nodes include a high concentration of non-core political and celebrity figures, such as \textit{Kangana Ranaut, Milind Soman, Satish Kaushik}, and international/foreign political figures such as \textit{Nawaz Sharif, Imran Khan, and Bilawal Bhuttozardari}, demonstrating a continued reliance on figures outside of national politics.

firstpost (FP) demonstrates a significant change in the distribution of influence between the two protests. In 2020-21, FP had a relatively flat distribution of eigenvector scores among the top 20 entities, with \textit{Narendra Modi} leading, but scores clustering together. FP also exhibits centralization of state political figures (e.g., \textit{Mamata Banerjee, Abhishek Banerjee, and Partha Chatterjee}, politicians from the state of West Bengal), in the political discourse in 2020-21.

Preliminary results in 2024 show that the distribution steepened considerably, led by \textit{Narendra Modi} and followed closely by \textit{Rahul Gandhi}. Additionally, we see inclusion of international political figures such as \textit{Emmanuel Macron} and \textit{Gurpatwant Singh Pannun} high in the list, reflecting their increased involvement within the core structural narrative. We also observe the presence of activists \textit{Umar Khalid, Sudha Bhardwaj}, and \textit{Varvara Rao} among the top 20 entities, indicating their co-occurrence with other highly influential entities in the discourse.

Across all four outlets in the 2020-21 period, only one farmer leader, \textit{Rakesh Tikait}, manages to appear among the top 20 most influential entities in terms of eigenvector centrality. This indicates that while farmer figures may have gained visibility (see appendix for degree centrality), they were largely unable to achieve the same level of structural influence held by eminent politicians in the network's core.

\begin{figure*}[!htbp]
    \centering
    \begin{subfigure}[b]{0.47\textwidth}
        \centering
        \includegraphics[width=\textwidth]{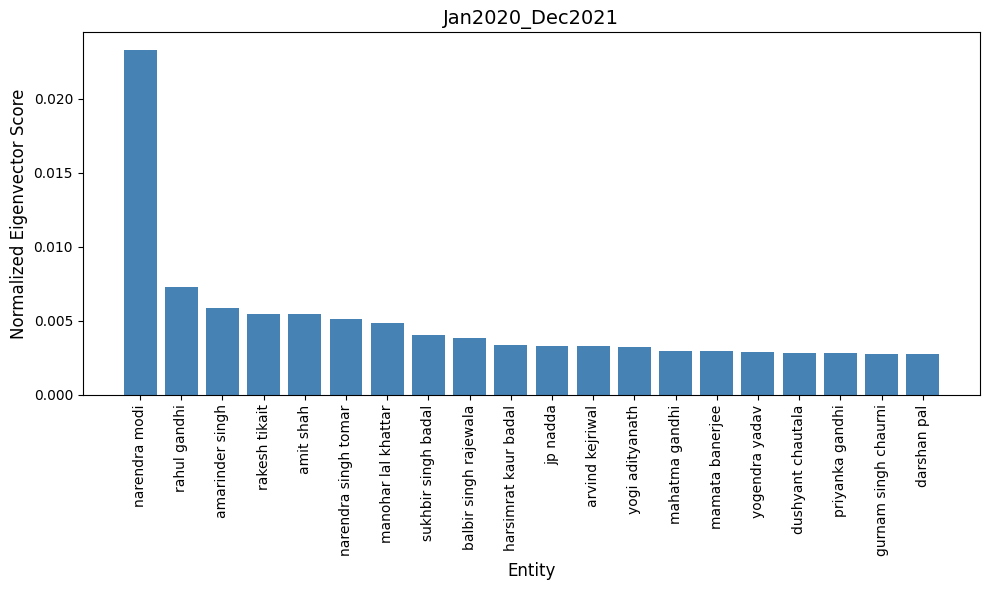}
        \caption{eigenvector of top 20 entities in 2020 (TOI)}
        \label{fig:eigen_2020_toi}
    \end{subfigure}
    \hfill
    \begin{subfigure}[b]{0.47\textwidth}
        \centering
        \includegraphics[width=\textwidth]{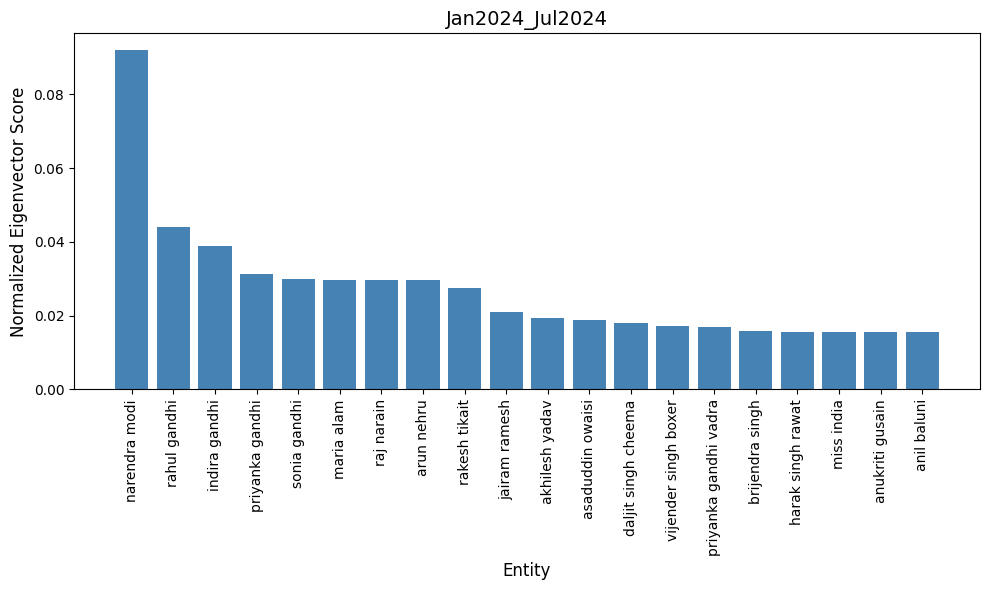}
        \caption{eigenvector of top 20 entities in 2024 (TOI)}
        \label{fig:eigen_2024_toi}
    \end{subfigure}
    \begin{subfigure}[b]{0.47\textwidth}
    \centering
    \includegraphics[width=\textwidth]{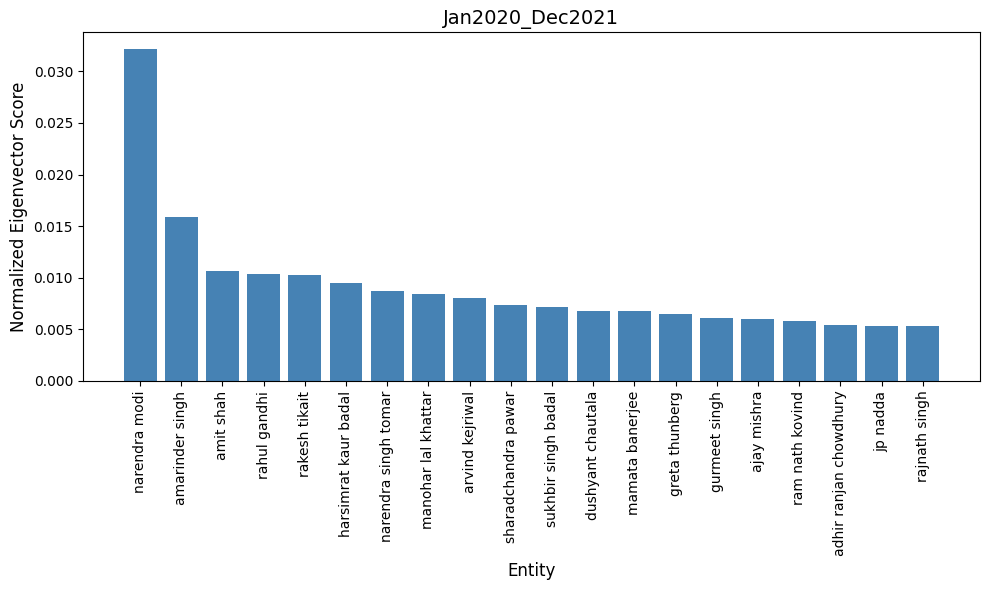}
        \caption{eigenvector of top 20 entities in 2020 (IE)}
        \label{fig:eigen_2020_ie}
    \end{subfigure}
    \hfill
    \begin{subfigure}[b]{0.47\textwidth}
        \centering
        \includegraphics[width=\textwidth]{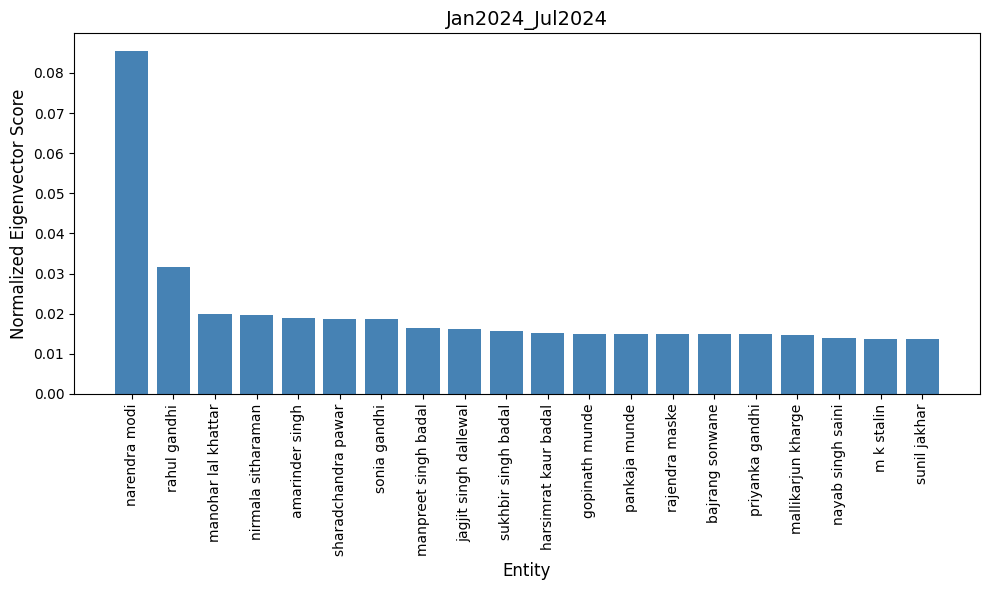}
        \caption{eigenvector of top 20 entities in 2024 (IE)}
        \label{fig:eigen_2024_ie}
    \end{subfigure}

    \begin{subfigure}[b]{0.47\textwidth}
    \centering
    \includegraphics[width=\textwidth]{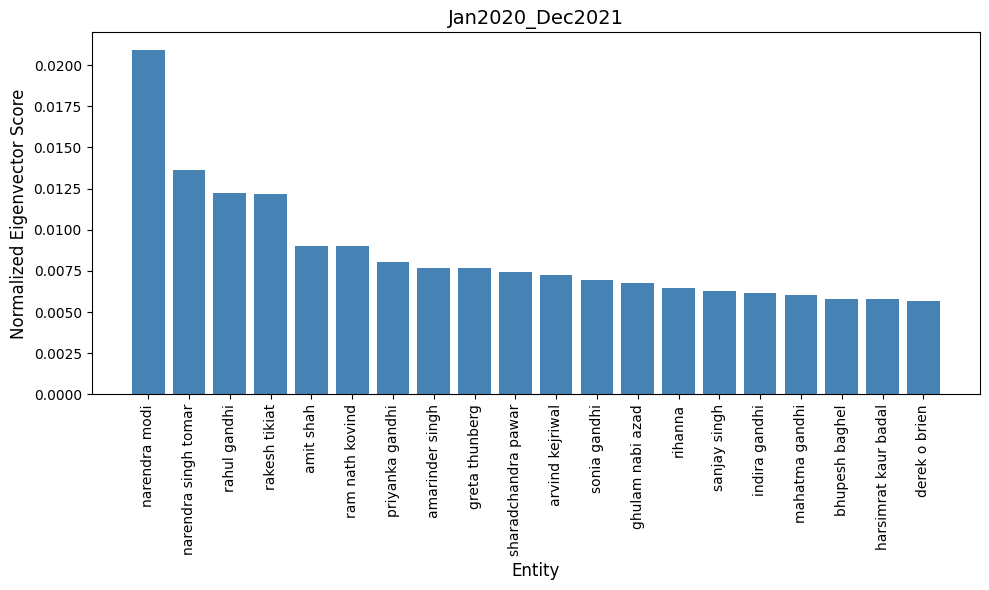}
        \caption{eigenvector of top 20 entities in 2020 (dna)}
        \label{fig:eigen_2020_dna}
    \end{subfigure}
    \hfill
    \begin{subfigure}[b]{0.47\textwidth}
        \centering
        \includegraphics[width=\textwidth]{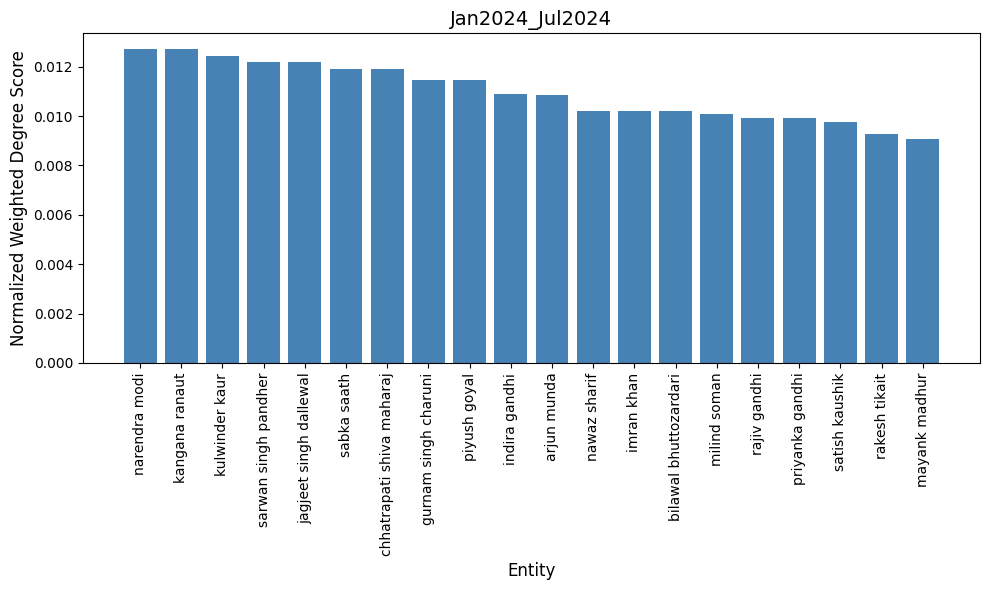}
        \caption{eigenvector of top 20 entities in 2024 (dna)}
        \label{fig:eigen_2024_dna}
    \end{subfigure}
    \begin{subfigure}[b]{0.47\textwidth}
    \centering
    \includegraphics[width=\textwidth]{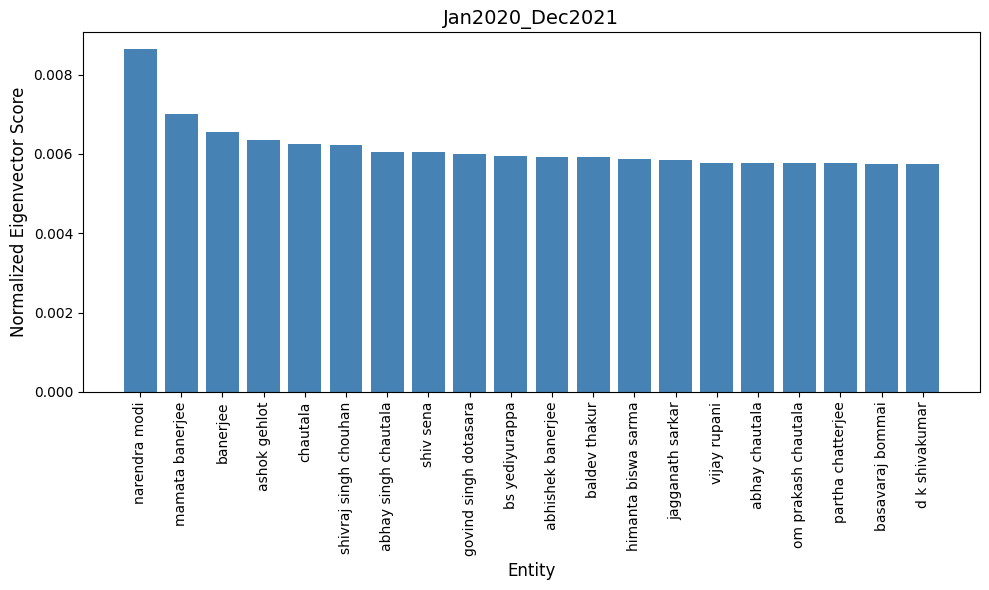}
        \caption{eigenvector of top 20 entities in 2020 (FP)}
        \label{fig:eigen_2020_FP}
    \end{subfigure}
    \hfill
    \begin{subfigure}[b]{0.47\textwidth}
        \centering
        \includegraphics[width=\textwidth]{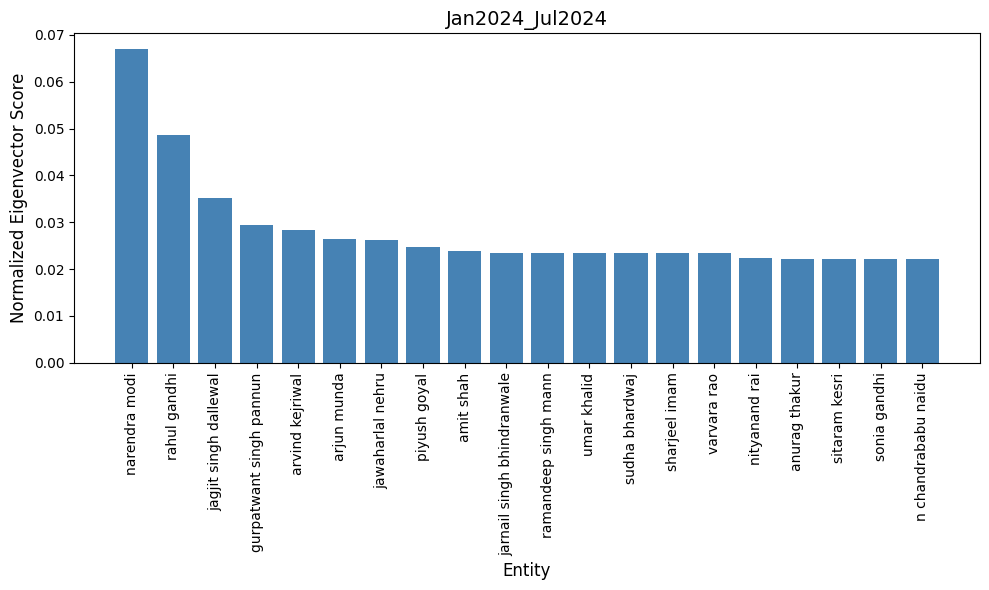}
        \caption{eigenvector of top 20 entities in 2024 (FP)}
        \label{fig:eigen_2024_FP}
    \end{subfigure}

    \caption{eigenvector centralities for Outlets}
    \label{fig:ev_comparison}
\end{figure*}

\subsection{betweenness Centrality}
The normalized betweenness Centrality analysis identifies entities that function as crucial brokers, lying on the shortest communication paths between different communities or segments of the political narrative. In all of the outlets considered for 2020-21, \textit{Narendra Modi} exhibits an exceptionally dominant role as the primary bridge node, with a score that is highly isolated from all subsequent entities. This suggests he was overwhelmingly mentioned with disparate elements of the network in news (e.g., government, regional leaders, opposition) during the initial protest period. We also see that for TOI, the top entities overwhelmingly comprise national politicians, unlike for the other sources where a mix of politician, sports-persons, and celebrities is observed. However, the general politician-centric discourse is evident across outlets.

Similar to figure \ref{fig:ev_comparison}, we find that across most outlets in 2020-21 (TOI, IE, and dna), the farmer leader \textit{Rakesh Tikait} ranks immediately among the top three or four most critical bridge nodes. This signifies his structural importance in connecting the localized farmer movement community to the broader national political and media structure. However, other than Tikait, few farmer leaders (except \textit{Gurnam Singh Charuni} for TOI and IE) appear in the media discourse for any news-source.

For both TOI and IE in 2024, while \textit{Narendra Modi} still leads, the high ranking of figures such as \textit{Kangana Ranaut}, \textit{Bhagwant Mann} (then state leader from the state of Punjab), and \textit{Rahul Gandhi} indicates that the discourse is more distributed among celebrity/political hybrids, opposition leaders, and key state leaders.

Preliminary results for dna in 2024 show the actor/politician \textit{Anupam Kher} achieving the highest score, followed by the high-profile opposition Chief Minister from the state of West Bengal \textit{Mamata Banerjee}. These findings hint towards the core narrative bridge for dna in 2024 being carried by figures known for cultural or regional opposition narratives.

In the 2024 FP plot based on limited data, the top two bridging roles are tightly held by \textit{Narendra Modi} and \textit{Mallikarjun Kharge} (a senior opposition leader). Additionally, \textit{Emmanuel Macron, Olaf Scholz} (international political figures) rank in this plot, suggesting that international political relationships played a substantial brokerage role in connecting different segments of the 2024 narrative in this outlet.

\begin{figure*}[!htbp]
    \centering
    \begin{subfigure}[b]{0.47\textwidth}
        \centering
        \includegraphics[width=\textwidth]{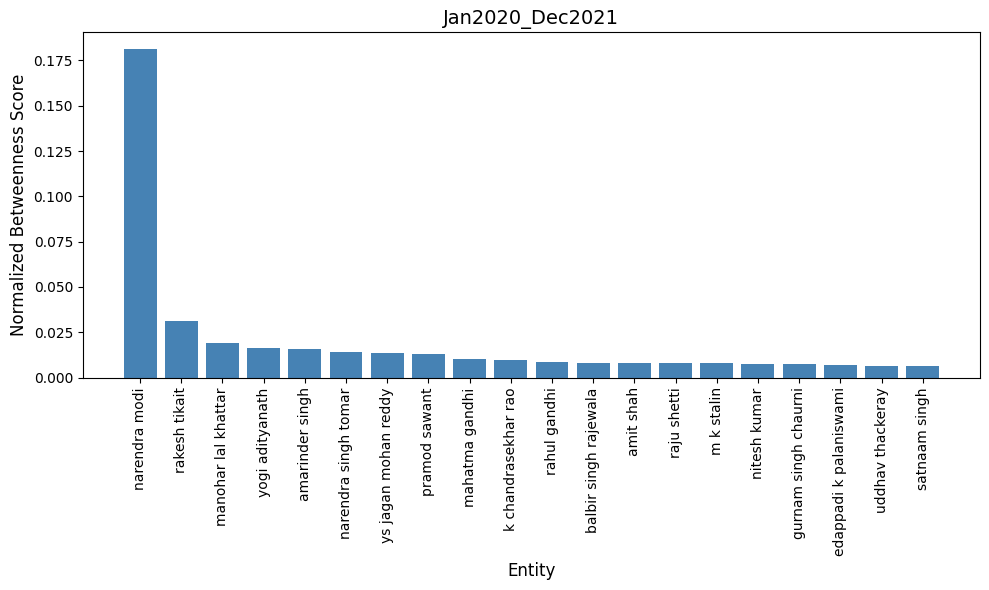}
        \caption{betweenness of top 20 entities in 2020 (TOI)}
        \label{fig:eigen_2020_toi}
    \end{subfigure}
    \hfill
    \begin{subfigure}[b]{0.47\textwidth}
        \centering
        \includegraphics[width=\textwidth]{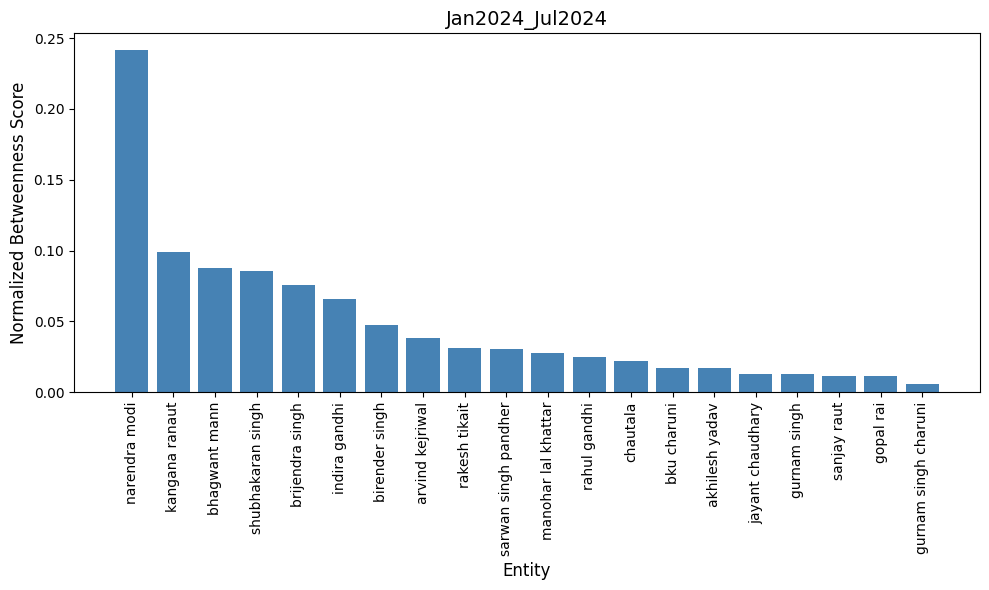}
        \caption{betweenness of top 20 entities in 2024 (TOI)}
        \label{fig:eigen_2024_toi}
    \end{subfigure}
    \begin{subfigure}[b]{0.47\textwidth}
    \centering
    \includegraphics[width=\textwidth]{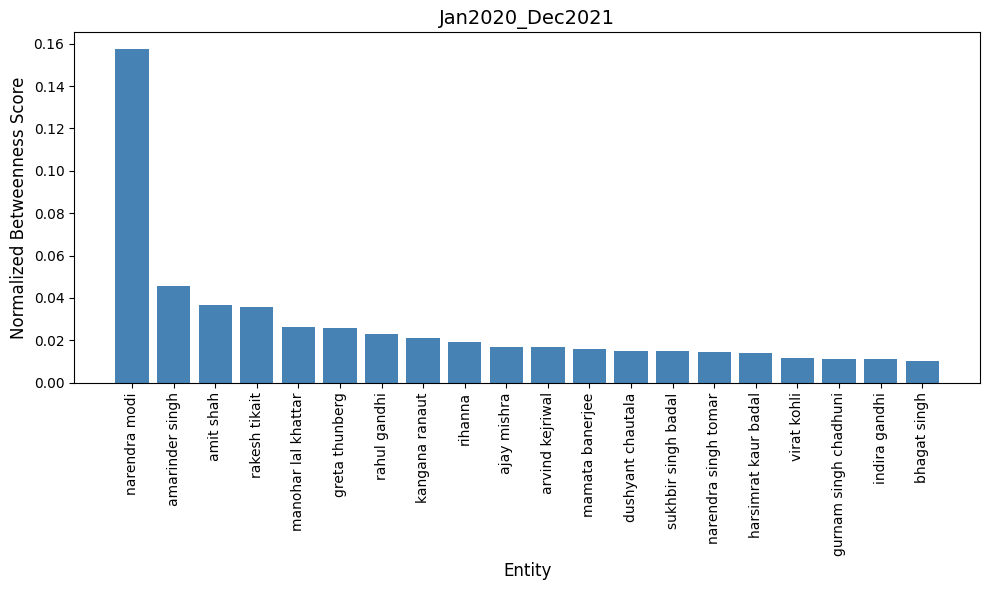}
        \caption{betweenness of top 20 entities in 2020 (IE)}
        \label{fig:eigen_2020_ie}
    \end{subfigure}
    \hfill
    \begin{subfigure}[b]{0.47\textwidth}
        \centering
        \includegraphics[width=\textwidth]{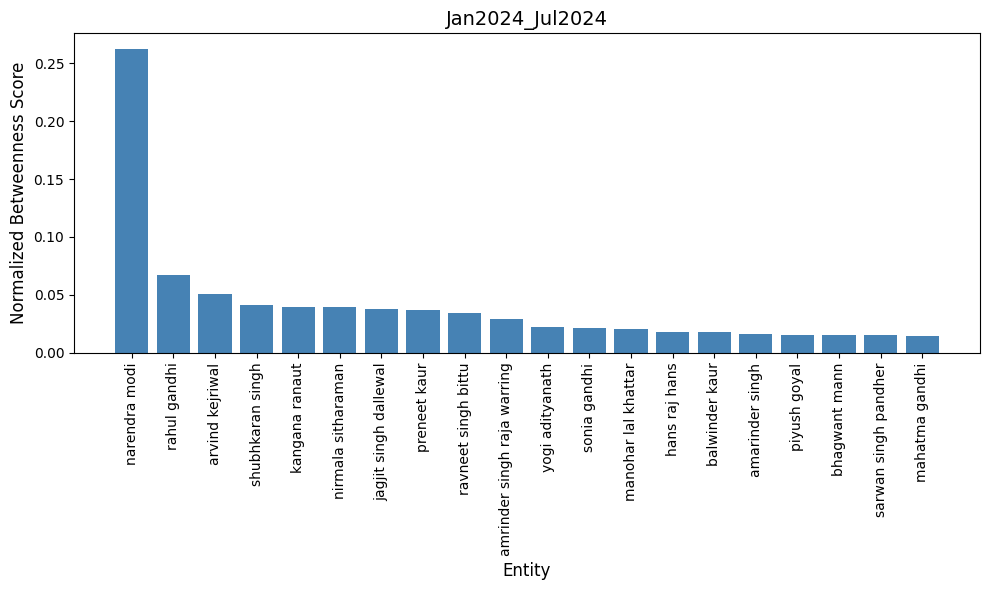}
        \caption{betweenness of top 20 entities in 2024 (IE)}
        \label{fig:eigen_2024_ie}
    \end{subfigure}

    \begin{subfigure}[b]{0.47\textwidth}
    \centering
    \includegraphics[width=\textwidth]{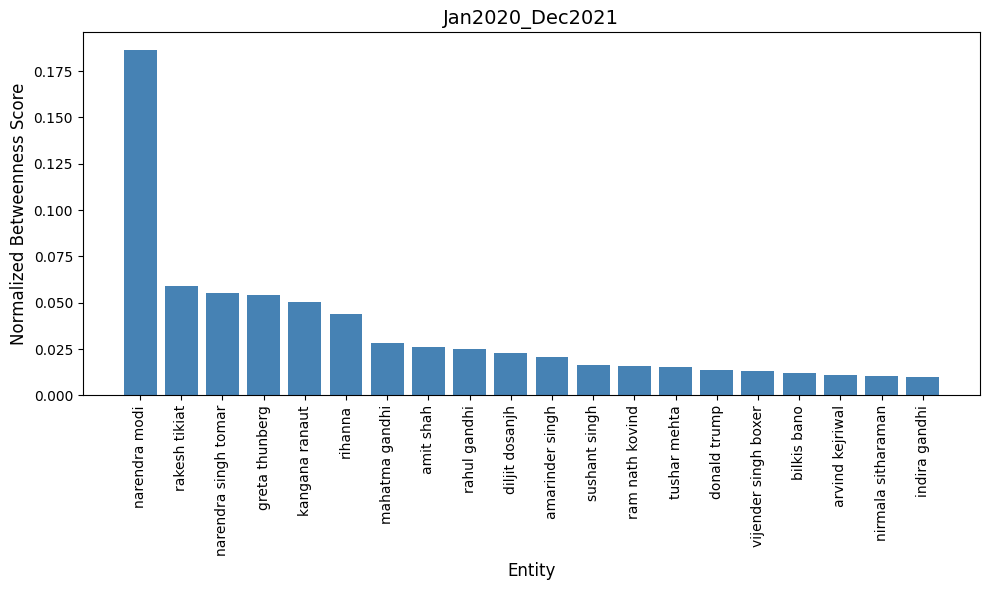}
        \caption{betweenness of top 20 entities in 2020 (dna)}
        \label{fig:eigen_2020_dna}
    \end{subfigure}
    \hfill
    \begin{subfigure}[b]{0.47\textwidth}
        \centering
        \includegraphics[width=\textwidth]{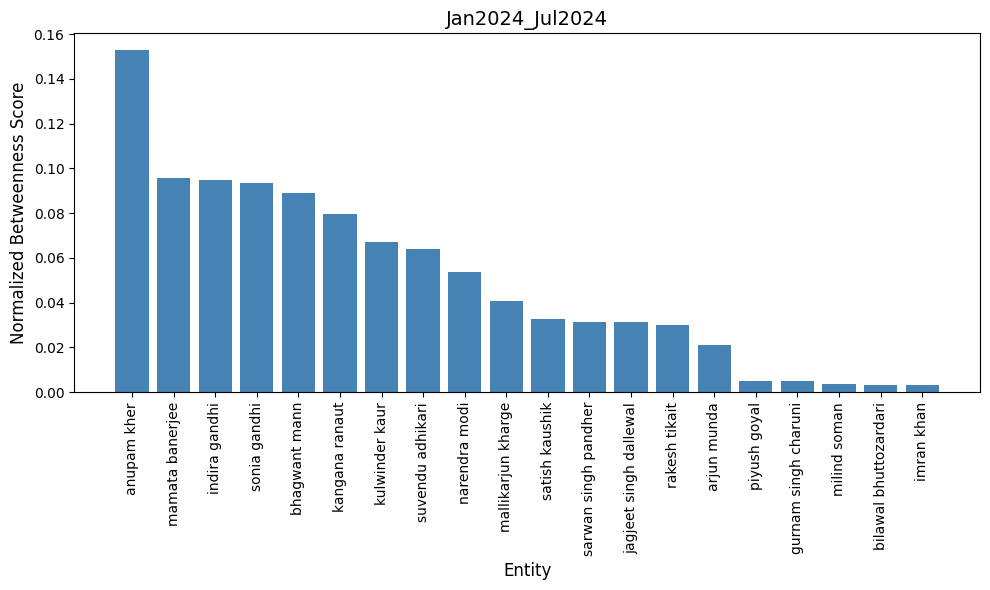}
        \caption{betweenness of top 20 entities in 2024 (dna)}
        \label{fig:eigen_2024_dna}
    \end{subfigure}
    \begin{subfigure}[b]{0.47\textwidth}
    \centering
    \includegraphics[width=\textwidth]{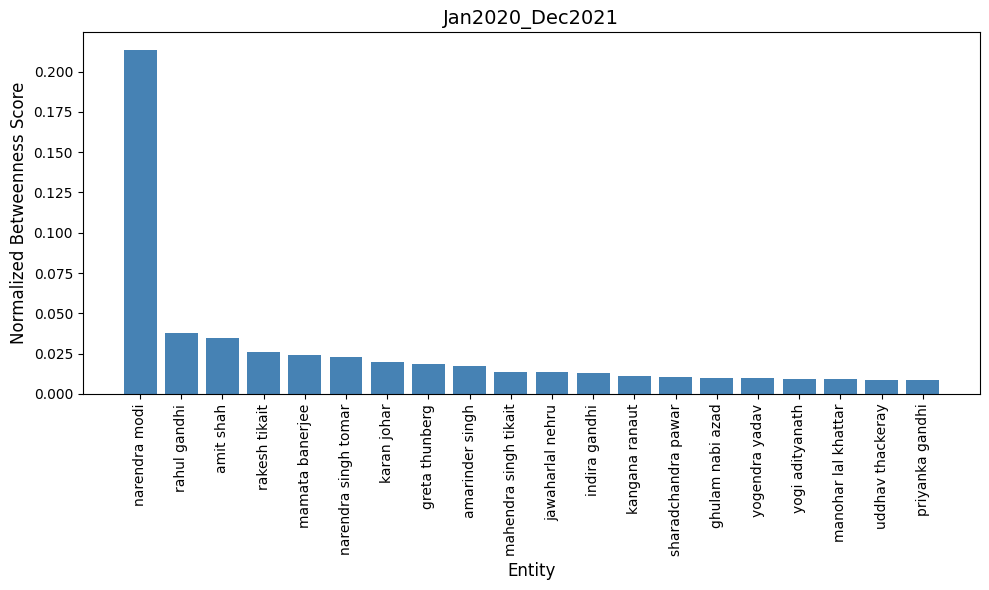}
        \caption{betweenness of top 20 entities in 2020 (FP)}
        \label{fig:eigen_2020_FP}
    \end{subfigure}
    \hfill
    \begin{subfigure}[b]{0.47\textwidth}
        \centering
        \includegraphics[width=\textwidth]{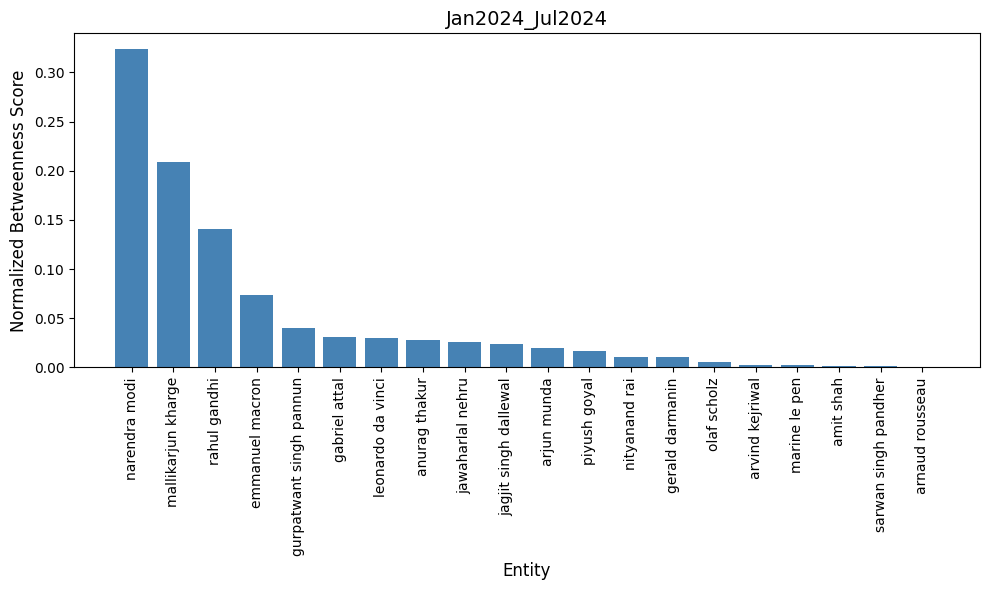}
        \caption{betweenness of top 20 entities in 2024 (FP)}
        \label{fig:eigen_2024_FP}
    \end{subfigure}

    \caption{betweenness centralities for Outlets}
    \label{fig:ev_comparison}
\end{figure*}

\subsection{Community Analysis}
To obtain an idea of the most prominent discourse communities in the Indian media, we attempted to perform community detection on MediaGraph using the \textit{Leiden} community detection algorithm. The selection of the Leiden community detection algorithm was motivated by its demonstrated stability in large-scale networks, coupled with its robustness and efficient execution. We ran the algorithm separately for the networks for each news-source, and for the two protest events. Next, we calculated the intra-community eigenvector centralities for the top three largest communities in terms of number of nodes. Our inferences in this section are thus based on the top five influential entities (with respect to their eigenvector centralities) of the three largest communities, for each news-source and each protest event (table \ref{tab:fullwidth}). As noted earlier, the findings for dna and FP in 2024 are constrained by their small network sizes, which limits the robustness of outlet-specific interpretations.

\renewcommand{\arraystretch}{1.15}    
\renewcommand{\tabularxcolumn}[1]{m{#1}} 

\begin{table*}[t]
    \centering 
    \caption{Top five community leaders (in terms of their eigenvector centrality) in the three largest communities for the two protests}
    \label{tab:fullwidth}

    \begin{tabularx}{0.95\textwidth}{|c|c|>{\centering\arraybackslash}X|>{\centering\arraybackslash}X|>{\centering\arraybackslash}X||>{\centering\arraybackslash}X|>{\centering\arraybackslash}X|>{\centering\arraybackslash}X|}
        \hline
        \multicolumn{2}{|c|}{} & \multicolumn{3}{c||}{\textbf{2020-21}} & \multicolumn{3}{c|}{\textbf{2024}} \\
        \hline
        & \textbf{Rank} & \textbf{Community 1} & \textbf{Community 2} & \textbf{Community 3} & \textbf{Community 1} & \textbf{Community 2} & \textbf{Community 3} \\
        \hline \hline

        \multirow{5}{*}{\textbf{TOI}} & 1 & Narendra Modi & Rakesh Tikait & Rahul Gandhi & Narendra Modi & Kangana Ranaut & Arvind Kejriwal \\ \cline{2-8}
                                      & 2 & Amit Shah & Balbir Singh Rajewala & Deepender Hooda & Rahul Gandhi & Sharmin Segal & Ravneet Singh Bittu \\ \cline{2-8}
                                      & 3 & JP Nadda & Yogendra Yadav & Priyanka Gandhi & Indira Gandhi & Ranbir Kapoor & Rajneesh Dhiman \\ \cline{2-8}
                                      & 4 & Yogi Adityanath & Deep Sidhu & Ajay Mishra Teni & Priyanka Gandhi & Sanjay Leela Bhansali & Vijay Rupani \\ \cline{2-8}
                                      & 5 & Narendra Singh Tomar & Lakha Sidhana & Raman Kashyap & Sonia Gandhi & Richa Chadha & Vijay Sampla \\
        \hline \hline

        \multirow{5}{*}{\textbf{IE}}  & 1 & Narendra Modi & Rajnath Singh & Amarinder Singh & Rahul Gandhi & Narendra Modi & Jagjit Singh Dallewal \\ \cline{2-8}
                                      & 2 & Sharadchandra Pawar & Narendra Singh Tomar & Ravneet Bittu & Nitish Kumar & Manohar Lal Khattar & Bhagwant Mann \\ \cline{2-8}
                                      & 3 & Rahul Gandhi & Amit Shah & Manish Tewari & Nitish Ji & Nayab Singh Saini & Piyush Goyal \\ \cline{2-8}
                                      & 4 & Ram Nath Kovind & Piyush Goyal & Preneet Kaur & Tejashwi Yadav & Rao Inderjit Singh & Nityananad Rai \\ \cline{2-8}
                                      & 5 & Sonia Gandhi & Nitin Gadkari & Jasbir Singh Gill & Rabri Devi & Krishan Pal Gurjar & Arjun Munda \\
        \hline \hline

        \multirow{5}{*}{\textbf{dna}} & 1 & Narendra Singh Tomar & Narendra Modi & Greta Thunberg & Sarwan Singh Pandher & Kangana Ranaut & Anupam Kher \\ \cline{2-8}
                                      & 2 & Rakesh Tikiat & Tanmanjeet Singh Desi & Rihanna & Jagjeet Singh Dallewal & Kulwinder Kaur & Santosh Singh \\ \cline{2-8}
                                      & 3 & Balbeer Singh Rajewala & Boris Johnson & Sachin Tendulkar & Gurnam Singh Charuni & Vishak Nair & Abhinandan Varthaman \\ \cline{2-8}
                                      & 4 & SA Bobde & Preet Kaur Gill & Virat Kohli & Piyush Goyal & Atal Bihari Vajpayee & Akanksha Singh \\ \cline{2-8}
                                      & 5 & Tushar Mehta & Jagmeet Singh & Pragyan Ojha & Arjun Munda & Jagjivan Ram & Ashish Vidyarthi \\
        \hline \hline

        \multirow{5}{*}{\textbf{FP}}  & 1 & Narendra Modi & Rakesh Tikait & Akshay Kumar & Emmanuel Macron & Narendra Modi & Rahul Gandhi \\ \cline{2-8}
                                      & 2 & Rahul Gandhi & Yogendra Yadav & Sunny Deol & Gabriel Attal & Jawaharlal Nehru & Gurpatwant Singh Pannun \\ \cline{2-8}
                                      & 3 & Narendra Singh Tomar & Darshan Pal & Ajay Devgn & Kevin Bertens & Sitaram Kesri & Melanie Joly \\ \cline{2-8}
                                      & 4 & Amit Shah & Balbeer Singh Rajewala & Diljit Dosanjh & Viktor Orbn & Sonia Gandhi & Mohammed Zubair \\ \cline{2-8}
                                      & 5 & Ram Nath Kovind & Gurnam Singh Chadhuni & Javed Akhtar & Ursula von der Leyen & N Chandrababu Naidu & Justin Trudeau \\
        \hline
    \end{tabularx}
\end{table*}

\subsubsection{Times of India}
We make the following observations for the 2020-21 communitities for TOI:
\begin{itemize}
    \item \textbf{Community 1 - Ruling Politicians:} This community clusters key political figures belonging to either the and central government or to the ruling party. These high prominence national politicians include \textit{Narendra Modi, Amit Shah, Yogi Adityanath,} and \textit{Narendra Singh Tomar}. 
    \item \textbf{Community 2 - Farmer Leaders}: This group represents the core farmers' movement in North India and consists of the prominent Farmer leaders like \textit{Rakesh Tikait and Balbeer Singh Rajewala}, alongside \textit{Aam Aadmi Party} leader \textit{Yogendra Yadav}, who was a primary voice in support of the protests. 
    \item \textbf{Community 3 - Opposition Politicians}: This cluster is led by opposition leaders (with the chief opposition party INC in the forefront) like \textit{Rahul Gandhi and Priyanka Gandhi,} and \textit{Deepender Singh Hooda}. 
\end{itemize}
The 2024 event network for TOI, on the other hand, exhibits important distinctions in terms of the communities:

\begin{itemize}
    \item \textbf{Community 1 - Popular Political Leaders} This cluster unites the current PM with opposition INC stalwarts. In fact, we see that only the PM finds presence as a community leader from the ruling party for the largest community. The remaining leaders are all influential opposition figures. 
    \item \textbf{Community 2: Celebrities} This community highlights mostly celebrities (\textit{Kangana Ranaut, Ranbir Kapoor, Richa Chadha}), especially those from film and entertainment networks, who engaged significantly more in the 2024 farmers’ movement. 
    \item \textbf{Community 3 - Second Tier Politicians} Here we see a farmer leader (\textit{Ravneet Singh Bittu}) alongside an opposition politician (\textit{Arvind Kejriwal}). The other three community leaders are second tier ruling politicians, who are either state level politicians (former CM of the state of Gujarat \textit{Vijay Rupani}) or chairmen of important government bodies (\textit{Vijay Sampla}). 
\end{itemize}
We thus see significant heterogeneity in the communities for 2024, in contrast to the more homogeneous communities of 2020-21. This finding is also indicative of the fact that in 2024, there might have been a lack of dedicated focus from media towards the actual protest developments, and we see coverage of a substantial number of entertainment personalities in the discourse. Thus, instead of a dedicated farmers' community as in 2020-21, there were generic political discussions where farmer leaders occasionally found significant presence. 

\subsubsection{Indian Express}
For IE, the 2020-21 communities exhibit the following broad characteristics:
\begin{itemize}
    \item \textbf{Community 1 - Popular Political Leaders}: This cluster features leaders across the national political spectrum, including both ruling and opposition leaders, with the PM at the forefront. 
    \item \textbf{Community 2 - Ruling Politicians}: Most individuals of this community again are political entities from the ruling party, occupying top government roles (e.g., the Home Minister \textit{Amit Shah}, and the agriculture minister \textit{Narendra Singh Tomar}, minister of commerce and industry \textit{Piyush Goyal} and minister of transport \textit{Nitin Gadkari}). They mostly form the decision making team that responded regularly to the farmers' demands. 
    \item \textbf{Community 3 - State Leaders}: This group highlights regional/state leadership in Punjab and Delhi, with a focus on prominent state politicians. We find a healthy mix of opposition politicians (like \textit{Manish Tewari, Jasbir Singh Gill}) and ruling politicians (\textit{Amarinder Singh, Preneet Kaur (spouse of Amarinder Singh)}, and \textit{Ravneet Bittu} among the community leaders.  
\end{itemize}
We again see important distinctions for the 2024 IE communities, when compared with the 2020-21 communities. 

\begin{itemize}
    \item \textbf{Community 1 - Opposition Leaders}: This community is led by the key opposition leader \textit{Rahul Gandhi}, with other important opposition leaders (\textit{Rabri Devi, Tejashwi Yadav}). However, ruling alliance members like \textit{Nitish Kumar} also find presence. Importantly, most of the community leaders here hail from the state of Bihar (a chiefly agrarian state). 
    \item \textbf{Community 2 - State Leaders}: Here, with the only presence of the PM, the community leaders consist of Haryana’s prominent political leaders from the ruling party (like CM of Haryana \textit{Nayab Singh Saini, Manohar Lal Khattar, Krishan Pal Gurjar}, and \textit{Rao Inderjit Singh}. It must be noted that Haryana was one of the states (alongside Punjab) from where the protests spawned initially, and exhibited prolonged sustenance\footnote{\textit{Singhu} and \textit{Tikri} borders in Haryana witnessed a year long protest in 2020-21, one of the largest protests in India in terms of scale, followed by a smaller, yet prolonged protest in 2024.}. 
    \item \textbf{Community 3 - Farmer Leaders and Ruling Politicians}: This community is led by a farmer leader (\textit{Jagjit Singh Dallewal}), followed by state leaders (CM of Punjab \textit{Bhagwant Singh Mann}) and central ministers from the ruling party(\textit{Piyush Goyal} and \textit{Arjun Munda} (union minister for tribal welfare)). 
\end{itemize}
Thus, for IE, we find that the exclusive political discourse in media in 2020 mostly sustained in 2024. However, when compared to TOI, IE provided greater coverage to regional political leaders (especially from the states of Punjab and Haryana) in 2024.

\subsubsection{dna}
For 2020-21, the dna discourse in the first protest exhibited three clearly separable communities:
\begin{itemize}
\item \textbf{Community 1: Ruling Politicians and Judiciary} This community was centralized around Government and Legal figures, led by \textit{Narendra Singh Tomar} and key farmer Leader \textit{Rakesh Tikait}, but also included the former Chief Justice of India (\textit{SA Bobde}) and the Solicitor General (\textit{Tushar Mehta}), indicating a focus on the legal and governmental machinery related to the farm laws.
\item \textbf{Community 2: Ruling and International Politicians}: This community clustered the PM closely with high-ranking International Political Figures from the UK and Canada (e.g., \textit{Boris Johnson, Tanmanjeet Singh Desi, Jagmeet Singh}), suggesting a narrative that linked the top domestic political authority with external diplomatic scrutiny.
\item \textbf{Community 3: Celebrities and Activists}: This community was a dedicated sphere of celebrities and international activists (\textit{Greta Thunberg, Rihanna, Sachin Tendulkar}), reflecting dna's focus on the cultural and global dimensions of the protest narrative.
\end{itemize}

In 2024, the limited data suggests that the structure became less compartmentalized and more sensationalized:
\begin{itemize}
    \item \textbf{Community 1: Ruling Politicians and Farmer Leaders} This community centered on farmer leaders (\textit{Sarwan Singh Pandher, Jagjeet Singh Dallewal, Gurnam Singh Charuni}), immediately followed by Union Ministers (\textit{Piyush Goyal, Arjun Munda}).
    \item \textbf{communities 2 and 3 - Celebrities and Historical Political Figures:} These communities are dominated by a unique and unusual mix of actors/public Figures (\textit{Kangana Ranaut, Kulwinder Kaur, Vishak Nair}), and historical/deceased political entities (\textit{Atal Bihari Vajpayee, Jagjivan Ram}). This shift suggests a shift from policy or legal structures towards actors known for cultural commentary, pro-government positions, or figures linked to specific, often sensational incidents. 
\end{itemize}
\subsubsection{firstpost}
We make the following observations for the 2020-21 communities for FP:
\begin{itemize}
\item \textbf{Community 1 - Ruling Politicians:} This community is centered around dominant national political figures, led by the Prime Minister (PM), with the presence of key ruling party members. The clustering suggests a narrative anchored in central political authority and decision-making, reflecting a top-down framing of the protest through the lens of national governance.

\item \textbf{Community 2 - Farmer Leaders:} This cluster consists of prominent farmer leaders such as \textit{Rakesh Tikait, Darshan Pal, Balbeer Singh Rajewala,}, and\textit{Gurnam Singh Charuni}. The grouping indicates a clearly identifiable farmer-led discourse, representing organizational leadership and mobilization efforts within the protest.

\item \textbf{Community 3 - Celebrities:} This community is dominated by film personalities and public figures (e.g., \textit{Akshay Kumar, Sunny Deol, Diljit Dosanjh, Javed Akhtar}), suggesting that FP devoted a distinct narrative space to celebrity engagement with the protest. This reflects an emphasis on the cultural and public-opinion dimensions of the movement.
\end{itemize}

In contrast, the limited data from 2024 event network for FP hints towards shifts in the composition and orientation of communities:
\begin{itemize}
\item \textbf{communities 1 and 3 - International Politicians:} These communities are led by international political figures such as \textit{Emmanuel Macron} and \textit{Gabriel Attal} and \textit{Viktor Orban} in community 1, and \textit{Melanie Joly} and \textit{Justin Trudeau} in community 3. Community 3 also shows presence of the the domestic opposition leader \textit{Rahul Gandhi}. Interestingly, the founder of the fake news combating media outlet \textit{Alt News} \textit{Mohammed Zubair} finds mention as a community leader in community 3. The prominence of such entities suggests that FP framed the 2024 protests within a broader international geopolitical context, linking domestic developments with global political discourse.

\item \textbf{Community 2 - Ruling and Historical Political Figures:} Leadership for this cluster includes the PM along with historical and ideological figures such as \textit{Jawaharlal Nehru} and \textit{Sitaram Kesri}, as well as politically contentious actors like \textit{N Chandrababu Naidu}. The co-occurrence of contemporary leaders with historical and activist figures indicates a narrative strategy that situates the protest within a longer political and ideological continuum.

\end{itemize}

Thus, while the 2020-21 FP communities exhibit a relatively structured separation between political leadership, farmer actors, and celebrity discourse, the 2024 communities hint towards a shift to internationalization of the issue. Notably, the absence of a distinct farmer-centric community in 2024 indicates a dilution of grassroots representation, with the discourse increasingly mediated through globally and politically symbolic actors.




\subsection{Link Prediction}
To get an idea about the predictability of co-occurrence links in MediaGraph, we evaluated the predictability of entity co-occurrence links through systematic experiments across TOI, IE, FP, and dna. These experiments tested four key dimensions: supervision paradigm, temporal sampling strategy, edge-weight thresholds, and inclusion of node-level structural properties. 

Owing to the limitations with the 2024 data and the necessity of large-scale data to train GraphSAGE, we restrict our analysis along the link predictability axis to 2020-21 protests. This ensures robustness of our findings along this axes.
We report the detailed results of all experiments in the Appendix.
\subsubsection{Overall Predictability (No Edge Weight Threshold)}
Table \ref{tab:allweights} shows the overall predictability of news-sources, in terms of the best prediction performance for all links included in the test set (no edge-weight thresholds applied). 
\begin{table*}[t]
\small
\centering
\caption{Best performing link prediction models across media outlets, for all links in the test data}
\label{tab:allweights}
\begin{tabular}{|p{2cm}|p{2.5cm}|p{2cm}|p{1.5cm}|p{2cm}|}
\hline
\textbf{Outlet} &
\textbf{Best Configuration} &
\textbf{Accuracy (\%)} &
\textbf{F1 Score} &
\textbf{Properties Included} \\ \hline

Times of India &
Supervised (One-time, Properties) &
71.02 & 0.7604 &
Community ID, degree, betweenness, eigen \\ \hline

Indian Express &
Supervised (One-time, Properties) &
71.82 & 0.7182 &
Community ID, degree, betweenness, eigen \\ \hline

dna &
Supervised (One-time, Properties) &
69.07 & 0.6332 &
Community ID \\ \hline

firstpost &
Supervised (One-time, Properties) &
80.08 & 0.7972 &
Community ID, degree \\ \hline

\end{tabular}
\end{table*}
We gain interesting insights from these findings. It can be seen that the news-sources have non-trivial predictability (with respect to baselines) when it comes to the task of link prediction (F1 score $\ge 0.63$ across sources), indicative of a predictable reporting style in terms of entity co-occurrences. We report the comparison of the GraphSAGE based link prediction results with that of two baseline models (random and community ID-based) in the Appendix.

\begin{table*}[t]
\small
\centering
\caption{Best performing link prediction models across media outlets, for links with edge weight threshold $>2$}
\label{tab:freqweights}
\begin{tabular}{|p{2cm}|p{2.5cm}|p{2cm}|p{1.5cm}|p{2cm}|}
\hline
\textbf{Outlet} &
\textbf{Best Configuration} &
\textbf{Accuracy (\%)} &
\textbf{F1 Score} &
\textbf{Properties} \\ \hline

Times of India &
Supervised (One-time, No Properties) &
95.17 & 0.9497 &
- \\ \hline

Indian Express &
Supervised (One-time, No Properties) &
91.82 & 0.9109 &
- \\ \hline

dna &
Supervised (One-time, No Properties) &
93.55 & 0.9394&
- \\ \hline

firstpost &
Supervised (One-time, No Properties) &
96.48 & 0.9647 &
- \\ \hline

\end{tabular}
\end{table*}

Additionally, without the application of any edge-weight threshold (i.e., all edges being considered in the test set), we find that the supervised approach with network properties (centralities and community IDs) as input significantly outperforms the unsupervised approach for all news-sources. Upon checking the correctly predicted links, we see that a majority of links exist between influential national politicians, which the model is easily able to predict based on its training data, dominantly consisting of similar links. This is indicative of a politician-heavy reporting style for most news-sources, leading to overwhelming presence of politician-politician links. 

\subsubsection{Effect of Edge-Weight Thresholds}
Our findings, however, change significantly when varying edge-weight thresholds are applied as can be seen in table \ref{tab:freqweights}. We apply a threshold $\ge 2$, which filters out any entity link occurring only once in the entire dataset. Firstly, we observe that the link prediction performance for all sources go up significantly with filtering out low weight links. In most cases, the increase in F1 score is around 30 percentage points when compared to the previous \textit{no threshold} setting. This improvement is partly understandable, since applying edge-weight thresholds restricts link prediction to only links with high weights (i.e., entity pairs occurring frequently), which reduces the size of the test data. The finding, however, also indicates the ability of the model to learn better about high frequency, obvious entity co-occurrences in the event (politician-politician links). Secondly, unlike the previous \textit{no threshold} setting, we find that in this case, the supervised approach without structural network properties performs the best for all news-sources.

\subsubsection{Effect of Structural Network Properties}
The network properties (Community ID and centralities) help capture the structural features of the network community to which an entity belongs. As can be seen from tables \ref{tab:allweights} and \ref{tab:freqweights}, these network properties especially help the model predict links for less frequent entity co-occurrences (no edge-weight threshold applied). These links are unobvious (e.g., links between less prominent politicians, or between politicians and less covered farmer leaders), and hence require the additional knowledge of network properties to be predicted. Taking the same properties as input also aids in improving the prediction for frequently occurring links (edge-weight threshold above 2), in some cases. However, for the edge-weight threshold setting, the best performance in all cases is obtained through the supervised approach devoid of network properties.

A probable cause of this phenomenon might be that for links with a threshold of $\geq 2$, the high frequency of such edges makes their prediction relatively straightforward for the model. In this regime, incorporating additional structural features may introduce noise or redundancy, thereby obscuring the underlying signal and leading to diminished performance. Nevertheless, it is important to note that even with this degradation, the supervised model augmented with structural properties still performs substantially better than in the \textit{no threshold} setting across all sources.
\subsubsection{Outlet-Specific Predictability}
For both settings (with and without edge-weight thresholding), we see a mostly consistent pattern of outlet level link predictability. The news-sources ranked in decreasing order of their F1 scores is FP, TOI, IE, and dna for the \textit{no threshold} setting, and FP, TOI, dna, and IE for the setting with edge weight threshold $\ge 2$. Thus, the fringe outlet FP, followed by the mainstream outlet TOI exhibit the highest link predictability in both settings. 

Additionally, we perform bootstrap resampling over the test edges to estimate confidence intervals (CI) and assess the statistical significance of differences in link predictability across sources. For the \textit{no threshold} setting, the results show that FP achieves the highest performance (95\% CI: [0.759, 0.805]), followed closely by TOI (95\% CI: [0.745, 0.788]), with only limited overlap in their confidence intervals, indicating consistently high predictability for both. In contrast, IE (95\% CI: [0.603, 0.711]) and dna (95\% CI: [0.556, 0.703]) exhibit substantially lower performance along with wider intervals, suggesting greater variability. Importantly, the confidence intervals for FP and TOI lie largely above those of IE and dna, providing strong evidence that differences in link predictability across sources are statistically robust and not attributable to sampling variability. We obtain similar findings for significance testing for the setting with thresholds applied. The significant differences in link predictability across sources are indicative of differential reporting of the same event, by different sources, in terms of the entities covered.

\section{Discussion}
We studied the entity reporting preferences of four Indian news-sources using their entity co-occurrence network across the two Farmers' Protest events, using the \textit{MediaGraph} framework. Our findings suggest that along the three axes of analysis, namely network centrality, network community, and link predictability, the news-sources vary significantly in their reporting preferences for the same event. While it will be overtly simplistic to connect these results with media bias, the variance in reporting preferences seen across news-sources for the three axes clearly hint at their potential to differentially skew readers' opinions on important socio-political events. Most of our empirical findings along the study of the three network theoretic axes corroborate the findings of previous literature around media analysis. In this section, we discuss our findings along the aforementioned axes of analysis.
\subsection{Network Centrality}
The analysis of top 20 entities for both Farmers' Protests in terms of their eigenvector centralities reveals interesting patterns. We see that while the mainstream outlets (TOI and IE) prefer a politician-heavy reporting of the protests, the fringe outlets (dna and FP) generally diversify their reporting trend by also carrying references to non-political and celebrity figures. Another motif of this diversity in fringe outlets is also their preference of covering influential non-central (state level) politicians. 

We also find that these patterns are consistent across both protest events (2020 and 2024). The fringe outlets retain their mix of political and non-political coverage in 2024, with more celebrity figures entering the foray. One of the primary drivers of this trend was a widely discussed altercation between politician-turned-celebrity \textit{Kangana Ranaut} and a security personnel, which drew substantial celebrity attention and amplified the discourse surrounding the second protest. Additionally, an interesting feature of FP centralizing political activists in 2024 is an indication of the outlet's tendency to link historical protest events with Farmers' Protests, an indication of indicating a broader pattern of narrative continuity and associative framing.

We can thus see that the fringe outlets have the tendency of diversify their reporting, by also co-mentioning non-political entities (celebrities, activists) alongside highly central political actors. This stands in contrast to mainstream outlets, indicating a deliberate attempt to broaden narrative scope and differentiate their coverage patterns. However, the conspicuous absence of farmer leaders among the most connected entities for all four sources (other than \textit{Rakesh Tikait} and \textit{Gurnam Singh Charuni}) is indicative of their broader marginalization of grassroots leadership, suggesting a shift in focus toward more politically salient and newsworthy/popular actors in the discourse. These trends paint a clear picture around reporting preferences of Indian digital print media: while most outlets focus on popular politicians and celebrities, there exists a clear lack of attention to structural issues and farmer leadership. Even when farmer leaders are mentioned, they are covered in a disconnected fashion, often alongside less prominent or weakly connected entities in the co-occurrence network. This has important implications for public discourse, as it risks diluting the visibility of core agrarian concerns and undermining the agency of farmer leadership. Over time, such framing may shape audience perceptions in ways that prioritize personalities over policy, potentially skewing both public understanding and policy attention.

\subsection{Community Analysis}
The community structures presented in Table 5 reveal not only differences across outlets, but also a substantive transformation in how protest discourse is organized between the 2020–21 and 2024 events. These shifts are particularly salient when viewed through the lens of community leadership, which serve as proxies for how media outlets conceptually group actors and frame the discourse.

In the 2020-21 protest, there is a consistent pattern across outlets (most clearly visible in TOI and FP) of functionally distinct and internally coherent communities, separating government actors, farmer leadership, and auxiliary domains such as opposition politics or celebrity engagement. For instance, in FP, we see a clear tripartite structure: national political leadership, farmer leaders, and a distinct celebrity cluster. This separation suggests that the protest was framed as a multi-stakeholder participation, where different actor groups retained identifiable narrative boundaries, indicative of relatively issue-centric reporting. 

Our finding of non-political actors being mostly absent in the centrality plots for mainstream outlets like TOI and IE (for both 2020 and 2024)  also holds in case of intra-community leadership. This suggests the propensity of mainstream sources to disproportionately cover political actor pairs in their coverage of the discourse. This finding corroborates the claims theorized in previous studies around the disproportionate coverage given by the mass media to the range and dynamics of governmental debates and popular political entities \cite{wl1990toward}.

In contrast, the 2024 protest exhibits a marked departure from this structure, particularly in fringe outlets such as FP and dna. FP’s communities in 2024 are no longer organized around domestic stakeholder groups; instead, they are dominated by international political figures (e.g., Emmanuel Macron), historical political actors, and ideologically salient or contentious figures (e.g., Gurpatwant Singh Pannun). dna too seems to follow this trend by covering celebrities (like actor turned politician Kangana Ranaut, Vishak Nair, Anupam Kher, and Ashish Vidyarthi) in two of its top three communuities. Notably, the absence of a dedicated farmer-leader community (other than a standalone presence of Gurnam Singh Charuni for dna) in 2024 represents a significant structural shift. Rather than being central to a cohesive cluster, farmer leaders are either absent from top community leadership or embedded within broader, less coherent political groupings. This finding is in line with previous studies around Farmers' Protests, which report relatively insignificant coverage of structural issues and entities around agriculture by Indian news media \cite{sen2019studying}.

This transition has important implications. First, it suggests a movement from actor-specific framing to narrative abstraction where the protest is no longer represented primarily through its immediate stakeholders, but through its connections to larger ideological or geopolitical entities. While such framing may indicate international attention, and enrichment of interpretation of the discourse, it simultaneously risks effacing the perspectives of primary actors, particularly farmer leadership, and structural issues of discussion from the media discourse.

Second, the increasing presence of heterogeneous communities, combining politicians, celebrities, and activists points to a broader trend of narrative hybridization. In 2020-21, celebrity engagement formed a distinct and bounded community (e.g., the third community for FP), whereas in 2024, such actors are either diffused across communities (as in dna) or replaced by historically and ideologically symbolic figures (like in dna and FP). This suggests a shift from peripheral amplification (celebrities reacting to protests) to integrative framing (embedding protests within larger symbolic narratives).

Finally, these structural differences highlight the role of editorial strategy in shaping discourse. Mainstream outlets such as TOI and IE maintain alignment majorly with political actors. This finding is in line with Traag et al. \cite{traag2016structure}, who report similar findings. In contrast, however, fringe outlets exhibit a stronger tendency toward covering non-political associations. Thus, the diversity in reporting preferences for fringe outlets as reported for centrality analysis is also corroborated at the community level. dna exclusively shows the present of judiciary entities in its discourse, establishing its status as an outlier in the source cohort.

\subsection{Link Predictability}
In terms of link predictability for the 2020-21 protest, we find that frequently co-occurring entity links (edge weight threshold above 2) are highly predictable for all sources ($F1>0.91$). This finding is indicative of the highly predictable reporting patterns of sources for prominent and obvious entity co-occurrences, which aligns with previous studies that empirically prove how media generally reinforces existing entity co-occurrences \cite{traag2016structure}. However, this performance declines when the edge-weight threshold is removed and the task shifts to predicting the full set of entity links. Under this setting, the performance of the GraphSAGE-based model deteriorates significantly ($F1 > 0.63$). This suggests that the sources also report a considerable number of low-frequency and non-obvious entity co-occurrences, which are inherently more difficult to predict and are less consistently represented in the learned network structure. However, even in the \textit{no threshold} setting, the link predictability for the sources is non-trivial, indicative of the predictable coverage of co-occurrences across all sources.

We also find that link predictability does not depend on an outlet being fringe or mainstream, for the four sources considered. We see that while TOI and IE show moderately high to high predictability, they do not gain the top spots. Surprisingly, FP comes across as the most predictable source in both settings (with and without edge weight threshold). Combining this finding with those around centrality and community analysis, we see that while the fringe outlets generally are more diverse in their entity coverage compared to mainstream sources, some of them (like FP) might be temporally more predictable in terms of the co-occurrences covered. This is not the case with dna, the other fringe outlet, which exhibits low predictability in both settings.

Combining our findings, we observe that although fringe outlets appear to diversify coverage at the level of individual entities, often incorporating a broader set of non-political actors, the structure of entity co-occurrences, particularly for FP, remains highly predictable. This reveals a deeper regularity beneath apparent diversity: while who is covered may vary, how entities are connected follows a consistent pattern. In this light, link predictability emerges as a powerful and previously underexplored dimension for assessing selection bias, shifting the focus from mere entity presence to the structure of relationships among them. 

\subsection{Misclassification Analysis}
To get an understanding of misclassifications done by the GraphSAGE based link predictor, we qualitatively checked the incorrect predictions by the model, for the setting where edge weight threshold was not applied (Table 6), since this setting generally resulted in the most number of misclassifications by the model. Two human experts, one from the author group and another outside it, were employed for the qualitative analysis. Both of the experts are knowledgeable around Indian politics and Farmers' Protests in general. The findings reported in this section have been reported based on three steps: (A) The experts first extracted the frequently occurring misclassification patterns from the test data, (B) They attempted to find the presence of the misclassified links in the training data, and in external sources (news articles outside the dataset, web-based resources, etc.), and (C) They analyzed the patterns around the misclassifications after three rounds of deliberation.

The misclassifications include the links that exist in the held-out test data, but were not predicted (false negatives), and the links that did not exist but were predicted by the model (false positives). We discuss the findings of our qualitative analysis of both of these misclassification types in this section.

\subsubsection{False Negatives}
The false negatives for dna reveal a clear and consistent pattern of cross-domain co-occurrences that are structurally weak but meaningful in the context of the 2020 Farmers' Protests. Specifically for dna, a large number of missed edges connect farmer leaders with mainstream political actors, such as \textit{Rakesh Tikait} with \textit{Yogi Adityanath}, and \textit{Gurnam Singh Charuni} with \textit{Sukhbir Singh Badal} and \textit{Gian Chand Gupta}. These links typically arise in infrequent, episodic contexts, such as negotiations, statements of support, or political criticism during the protests. Similarly, intra-opposition and intra-party associations (e.g., between \textit{Priyanka Gandhi and Bhupesh Baghel} are often embedded within broader group narratives, reducing the prominence of these specific ties. Additionally, the presence of low-salience or localized actors (e.g., regional leaders and lesser-known individuals like \textit{Dola Sen} and \textit{Ajay Kumar Lallu}) further contributes to sparsity, making such edges difficult for the model to learn despite their real-world relevance.

For FP, the misclassifications are also strongly indicative of heterogeneity in entity associations. A substantial portion of missed links involves pairings between politicians, historical figures, international personalities, and celebrities, e.g. between \textit{Asaduddin Owaisi} (opposition politician) with \textit{Rihanna} and \textit{Greta Thunberg}, and \textit{Kangana Ranaut} with \textit{Munawar Faruqui} (Indian comedian and rapper). These co-occurrences reflect broader media discourse during the protests, where international endorsements, celebrity commentary, and ideological debates were invoked. However, most of this co-occurrences occurred in bursts, leading to their insignificance when compared to the dominant political links. Moreover, FP exhibits a notable tendency to link contemporary actors with historical or ideological figures (e.g., \textit{Syama Prasad Mookerjee, Bhagat Singh}), indicating a framing strategy that situates current events within a larger political-historical narrative. Such associations are inherently irregular, often appearing in opinion pieces or contextual analyses rather than routine reporting, which explains their weak structural footprint. Consistent with dna, FP also consists of several links between regional political leaders (\textit{Deepender Hooda, Umesh Malik, Dharmendra Malik}), which again are misclassified owing to their episodic occurrences. 

For TOI, the misclassified links reveal a narrative that blends mainstream electoral friction, corporate antagonism, and deep-seated cultural symbolism. Similar to dna and FP, the model struggles with episodic and cross-domain associations, particularly those involving the intersection of the protest with the Lakhimpur Kheri violence \cite{lakhimpur2021}, e.g., the missed links between \textit{Ajay Mishra} (Member of Parliament from Kheri constituency in Uttar Pradesh) and \textit{Lovepreet Singh} (a sportsperson and protestor who died during the protest)/\textit{Shiv Kumar} (a labour rights activist and protestor). TOI also shows a significant number of missed edges regarding intra-party volatility, specifically within the Congress party’s Punjab and Uttarakhand units (e.g., \textit{Rahul Gandhi} with \textit{Navjot Singh Sidhu} and \textit{Harish Rawat}) where the protests served as a backdrop for leadership crises. A unique facet in the TOI data is the frequent co-occurrence of corporate figures with political agitators, such as \textit{Mukesh Ambani } linked with \textit{Navjot Singh Sidhu} or \textit{Sukhjinder Singh Randhawa}. This reflects the movement’s "anti-corporate" rhetoric, which appeared in news cycles primarily as specific accusations rather than stable structural relationships. Furthermore, consistent with FP’s historical framing, TOI contains missed links between contemporary actors and historical religious icons like \textit{Guru Nanak} and \textit{Guru Gobind Singh}, indicating that the media captured the spiritual and identity-based mobilization of the protesters in specialized reporting. The model, however, likely viewed these links as ``noise" when compared to standard political reporting.

For IE, the misclassified links underscore the publication’s focus on high-level political maneuvering, alliance-building, and the ideological-corporate nexus. Consistent with dna and TOI, the model fails to capture intra-party friction and leadership transitions, particularly within the Congress (e.g., \textit{Manish Tewari} with \textit{Sonia Gandhi}/\textit{Navjot Singh Sidhu}) and the \textit{Shiromani Akali Dal} (e.g., \textit{Sukhbir Badal} with \textit{Bikram Majithia}). However, IE displays a unique concentration of missed edges involving strategic electoral realignments (e.g., \textit{Akhilesh Yadav} with \textit{Jayant Chaudhary}/\textit{Priyanka Gandhi}) and Gujarat (e.g., \textit{Hardik Patel} with \textit{Jignesh Mevani}). Similar to FP and TOI, IE frequently links contemporary actors with historical and ideological pillars (e.g., \textit{BR Ambedkar} and \textit{Mahatma Gandhi}), reflecting a narrative style that frames the Farmers' Protests as a struggle for democratic values and constitutional rights. Similar to TOI, connections between state executive heads, ideological leaders, and corporate interests are at times missed (e.g., \textit{Manohar Lal Khattar} with \textit{Mohan Bhagwat} and \textit{Mukesh Ambani}), suggesting that the model struggles with IE’s tendency to report on the "behind-the-scenes" machinery of governance and its intersections with the protest.

The overall findings across all four newspapers suggesst that misclassifications often occur for links that consistently correspond to episodic, cross-domain, and structurally weak/infrequent co-occurrences rather than stable, frequently observed political associations. For dna, a distinct, yet salient pattern is the systematic misclassification of farmer–politician links, reflecting the model’s inability to capture interaction-driven but infrequent protest dynamics as reported in the source.

\subsubsection{False Positives}
The false positives predicted by the GraphSAGE model for dna largely reflect structural generalization and embedding proximity rather than actual co-occurrences. Many of these predicted links arise because the model learns community-level similarity. For instance, actors such as \textit{Rakesh Tikait, Yogendra Yadav,} and \textit{Darshan Pal} occupy similar neighborhoods in the protest network, leading the model to infer plausible but non-existent links with other politically or socially adjacent figures. Similarly, high-centrality political actors like \textit{Narendra Modi, Amit Shah,} and \textit{Rahul Gandhi} are connected to a wide range of entities in training, causing the model to over-generalize and predict links with loosely related or contextually co-mentioned actors (e.g., celebrities or international figures such as \textit{Rihanna}). Another important factor is the presence of latent thematic or discourse-level similarity, where entities are connected through shared narratives rather than explicit co-occurrence. For example, anti-corruption activism (linking \textit{Anna Hazare} with protest figures). The model also appears sensitive to coarse-grained type similarity (politician-politician, celebrity-celebrity), leading to spurious predictions like \textit{Vir Das} with \textit{Deepika Padukone}.

For FP, the false positive predictions indicate that the GraphSAGE model tends to over-generalize from loose co-occurrence patterns. Many predicted links connect mainstream political actors (e.g., \textit{Narendra Modi, Arvind Kejriwal, Amit Shah}) with celebrities, international figures, or cultural personalities (e.g., \textit{Rihanna, Bill Murray}), reflecting FP’s style of blending political reporting with global commentary. The model appears to interpret repeated co-mentions, especially during the Farmers’ Protests when international and celebrity attention was high, as evidence of structural relationships, even when no stable link exists.

Additionally, several false positives involve farmer leaders and regional actors (e.g., \textit{Yogendra Yadav, Gurnam Singh Charuni, Balbeer Singh Rajewal}) being linked to a wide range of unrelated entities due to dense co-mention within protest coverage. The model also exhibits semantic drift, incorrectly inferring links between contemporary actors and historical or symbolic figures (e.g., \textit{Mahatma Gandhi, Guru Tegh Bahadur}). Overall, these errors suggest that while the model captures co-occurrence signals well, it struggles to distinguish meaningful interactions from incidental narrative proximity.

For TOI, the false positives occur likely due to contextual co-occurrence and neighborhood overlap in news reporting. For instance, the link between \textit{Narendra Singh Tomar} and \textit{Santokh Singh Chaudhary} (a Congress MP who supported the protests) was possibly predicted incorrectly, since they both shared high-frequency connections with common entities around the protests. Similarly, the pairing of \textit{Rakesh Tikait} with \textit{Sudhanshu Trivedi} or \textit{Madan Kaushik} stems from the adversarial yet common co-occurrence neighborhood of farmer leaders and relatively low influence ruling politicians.

The model also struggles with semantic similarity among high-influence nodes. Many of these false positives involve opposition figures from different regions who were unified only by their stance against the central government, such as \textit{Mamata Banerjee} and \textit{Raghav Chadha}, or \textit{Sukhbir Singh Badal} and \textit{MK Stalin}. Because these individuals all shared a similar feature neighborhood (e.g., criticizing the same bills, appearing in ``Opposition Unity" articles), GraphSAGE inferred a direct link. Furthermore, the inclusion of historical figures like \textit{Mahatma Gandhi} alongside modern activists like \textit{Phulchand Tirkey} or \textit{Medha Patkar} highlights the model’s tendency to predict the rhetorical analogies used in Indian journalism, where current protests are frequently compared to historical revolts.

The false positives appearing in the IE dataset reflect the publication’s signature focus on policy-heavy reporting, national security narratives, and cultural commentary. Unlike TOI data, where the model often predicted rapid-fire adversarial connections, the IE false positives indicate the model's struggle with IE’s tendency to synthesize domestic politics with broader statecraft and global diplomacy. For instance, the pairing of \textit{Raghav Chadha} and National Security Advisor \textit{Ajit Doval} stems from IE's in-depth coverage of the intersection of political dissent with national security. The model also captures national-global connections, as seen in the link between \textit{Narendra Modi} and \textit{Benjamin Netanyahu}. Once again, the common errors around connecting opposition political actors with anti-corruption activists (\textit{Harsimrat Kaur Badal} and \textit{Anna Hazare}), and celebrity-celebrity connections (\textit{Virat Kohli} and \textit{Richa Chadha}) exist in case of IE.

Overall, across all sources, false positives stem from the model over-relying on co-occurrence and neighborhood similarity, leading to links between actors who share narrative context but lack real interactions. Common errors include connecting high-centrality politicians, protest leaders, and cross-domain entities (e.g., celebrities, international or historical figures). Among differences, dna mostly reflects structural proximity in protest networks, FP amplifies cross-domain discourse, TOI captures adversarial/electoral co-occurrence, and Indian Express reflects policy and global framing.


\subsection{Limitations and Future Work}
A primary limitation of this study is the analysis being restricted to English language news-sources. A majority of Indian news readers consume news from regional sources. As part of future work, we intend to include regional sources in multiple languages in the study, along with more English sources, to further generalize our findings.

We used Media Cloud, a publicly available news repository, to collect news articles from various sources. Provided the size of our dataset for the 2020-21 protests, the findings of this study are representative. However, our findings around the fringe outlets for the 2024 protests are not robust, and subject to deeper analysis with data augmentation. Additionally, our data collection process does not ensure completeness of the data collected, since Media Cloud does not guarantee a complete coverage of articles across time. The Media Cloud API also sometimes provided erroneous article URLs that could not be accessed by the crawler during the data collection phase. Although several news-sources in India do not possess well maintained and regularly updated article archives, and alternative data sources like GDELT also suffer from similar issues around data completeness, we plan to enrich our article collection by augmenting data from these other sources.

Our Entity Resolution heuristic is currently rule-based, and has multiple scopes of improvement. While its performance on the current dataset is above par, we plan to enrich it through representation learning, wherein the context around entities mentioned is taken into consideration. Although errors in the entity resolution pipeline are minimal, their potential propagation necessitates cautious interpretation of the results.m 

We have qualitatively checked the correctness of article data collected, by studying a sample of articles for each news-source and checking their relevance. However, there still lies scope for inaccuracies, albeit not significant, which may slightly impact the findings of this paper. Developing a transformer-based relevance classifier that classifies articles into relevant and irrelevant may help further in this context. 

An interesting direction for future work might also be the study of the latent factors that impact link predictability. Media ownership, business interests, and general political pressure may lead to over- or under-coverage of certain entity links, leading to a selection and coverage bias around them. Inclusion of more mainstream and fringe media outlets may also help us in establishing robustness of the current findings around link predictability.

\section{Conclusion}
This work presents \textit{MediaGraph}, a network theoretic framework to examine news reporting preferences across four Indian news outlets during the 2020-21 and 2024 Farmers’ Protests. By jointly analyzing centrality, community structure, and link predictability, we show that outlets differ not only in which entities they cover, but also in how they organize relationships or linkages among them. Mainstream outlets exhibit a consistent focus on prominent politicians, whereas fringe outlets introduce greater surface-level diversity by incorporating celebrities, activists, and regional figures. However, across sources, farmer leadership remains covered insignificantly, pointing to a broader marginalization of grassroots actors and structural issues in media discourse. Community-level analysis further reveals a temporal shift from relatively coherent, stakeholder-driven communities in 2020-21 to more heterogeneous groupings in 2024, particularly in fringe outlets, where discourse becomes increasingly entangled with international, historical, and symbolic narratives.

Our GraphSAGE-based link prediction experiments reveal that all of the news-sources considered in this study exhibit non-trivial link predictability of entity co-occurrences, indicative of a predictable reporting preference of selecting common entity links between prominent entities. Our link prediction experiments also uncover an important distinction between diversity and regularity in reporting. While fringe outlets appear more diverse in terms of entity coverage, certain fringe sources like firstpost exhibit highly predictable patterns in entity co-occurrences, indicating stable underlying reporting preferences. This suggests that apparent diversity in entity coverage does not necessarily translate to structural unpredictability around entity co-occurrences. Therefore, this study posits link predictability as a novel and powerful lens for empirically media bias, extending beyond entity presence to the patterns of association between entities that shape public discourse. Our overall findings highlight how editorial strategies manifest not only in entity coverage choices but also in the relational structure around entities, with important implications for how audiences perceive and interpret complex socio-political events.
\bibliographystyle{ACM-Reference-Format}
\bibliography{sample_base}

@String{Computing = "Computing" }

@String{Springer = "Springer-Verlag" }

@article{budak2016fair,
  title={Fair and balanced? Quantifying media bias through crowdsourced content analysis},
  author={Budak, Ceren and Goel, Sharad and Rao, Justin M},
  journal={Public Opinion Quarterly},
  volume={80},
  number={S1},
  pages={250--271},
  year={2016},
  publisher={Oxford University Press US}
}

@article{mccombs1972agenda,
  title={The agenda-setting function of mass media},
  author={McCombs, Maxwell E and Shaw, Donald L},
  journal={Public opinion quarterly},
  volume={36},
  number={2},
  pages={176--187},
  year={1972},
  publisher={Oxford University Press}
}

@article{djankov2003owns,
  title={Who owns the media?},
  author={Djankov, Simeon and McLiesh, Caralee and Nenova, Tatiana and Shleifer, Andrei},
  journal={The Journal of Law and Economics},
  volume={46},
  number={2},
  pages={341--382},
  year={2003},
  publisher={The University of Chicago Press}
}

@article{haryanto2011media,
  title={Media ownership and its implications for journalists and journalism in Indonesia},
  author={Haryanto, Ignatius},
  journal={Politics and the media in twenty-first century Indonesia: Decade of democracy},
  pages={104--118},
  year={2011},
  publisher={Routledge London, UK}
}

@inproceedings{wang2025media,
  title={Media bias detector: Designing and implementing a tool for real-time selection and framing bias analysis in news coverage},
  author={Wang, Jenny S and Haider, Samar and Tohidi, Amir and Gupta, Anushkaa and Zhang, Yuxuan and Callison-Burch, Chris and Rothschild, David and Watts, Duncan J},
  booktitle={Proceedings of the 2025 CHI Conference on Human Factors in Computing Systems},
  pages={1--27},
  year={2025}
}

@article{spinde2021automated,
  title={Automated identification of bias inducing words in news articles using linguistic and context-oriented features},
  author={Spinde, Timo and Rudnitckaia, Lada and Mitrovi{\'c}, Jelena and Hamborg, Felix and Granitzer, Michael and Gipp, Bela and Donnay, Karsten},
  journal={Information Processing \& Management},
  volume={58},
  number={3},
  pages={102505},
  year={2021},
  publisher={Elsevier}
}

@misc{fp1,
  author       = {{Wikipedia contributors}},
  title        = {2020--2021 Indian farmers' protest},
  year         = {2026},
  url = {https://en.wikipedia.org/wiki/2020%E2%80%932021_Indian_farmers%27_protest},
  note         = {Accessed: 2026-04-02}
}

@misc{fp2,
  author       = {{Wikipedia contributors}},
  title        = {2024--2025 Indian farmers' protest},
  year         = {2026},
  url = {https://en.wikipedia.org/wiki/2024%E2%80%942025_Indian_farmers%27_protest},
  note         = {Accessed: 2026-04-02}
}

@inproceedings{sen2019studying,
  title={Studying the discourse on economic policies in India using mass media, social media, and the parliamentary question hour data},
  author={Sen, Anirban and Ghatak, Debanjan and Kumar, Kapil and Khanuja, Gurjeet and Bansal, Deepak and Gupta, Mehak and Rekha, Kumari and Bhogale, Saloni and Trivedi, Priyamvada and Seth, Aaditeshwar},
  booktitle={Proceedings of the 2nd ACM SIGCAS Conference on Computing and Sustainable Societies},
  pages={234--247},
  year={2019}
}

@article{hosseinmardi2025unpacking,
  title={Unpacking media bias in the growing divide between cable and network news},
  author={Hosseinmardi, Homa and Wolken, Samuel and Rothschild, David M and Watts, Duncan J},
  journal={Scientific Reports},
  volume={15},
  number={1},
  pages={17607},
  year={2025},
  publisher={Nature Publishing Group UK London}
}

@inproceedings{horych2025promises,
  title={The promises and pitfalls of LLM annotations in dataset labeling: A case study on media bias detection},
  author={Horych, Tom{\'a}{\v{s}} and Mandl, Christoph and Ruas, Terry and Greiner-Petter, Andr{\'e} and Gipp, Bela and Aizawa, Akiko and Spinde, Timo},
  booktitle={Findings of the Association for Computational Linguistics: NAACL 2025},
  pages={1370--1386},
  year={2025}
}

@article{daoudi2026media,
  title={Media bias in sports journalism: A comparative study of Qatar 2022 World Cup coverage},
  author={Daoudi, Omar and Gainous, Jason and Hussain, Syed Ali and Zamoum, Khaled},
  journal={Communication \& Sport},
  volume={14},
  number={1},
  pages={207--230},
  year={2026},
  publisher={SAGE Publications Sage CA: Los Angeles, CA}
}

@article{liu2026war,
  title={The War of Ideas: Institutions and Global Media Bias},
  author={Liu, Sibo and Makarin, Alexey and Wu, Jinfeng and Zhang, Dong},
  year={2026},
  publisher={CESifo Working Paper}
}

@article{shahid2025mapping,
  title={Mapping Media Bias: Global Islamophobic Trends and their Reflections in South Asia},
  author={Shahid, Nidaa and Ghazanfar, Bilal},
  journal={The Beacon Journal},
  volume={5},
  number={02},
  year={2025}
}

@article{agustian2025analyzing,
  title={Analyzing Media Bias in Support of Government Policies: A Critical Discourse Analysis in a Newspaper},
  author={Agustian, Agung Farid},
  journal={PAROLE: Journal of Linguistics and Education},
  volume={15},
  number={1},
  pages={38--46},
  year={2025},
  publisher={Master Program in Linguistics, Diponegoro University}
}

@article{yogevmeasuring,
  title={Measuring Media Bias Toward Reform Prosecutors: A Multi-Method NLP Analysis of Bay Area News Coverage, 2019--2024},
  author={Yogev, Dvir and Law, Criminal and Center, Justice}
}

@article{belghoul2025media,
  title={Media Bias in Reporting the 2021 Sheikh Jarrah Evictions: A Van Dijkian Discourse Analysis},
  author={Belghoul, Hadjer and Baraka, Abdellah},
  journal={Majallat al-Nas}, 
  volume={12},
  number={2},
  pages={592--608},
  year={2025},
  publisher={ASJP}
}

@misc{lakhimpur2021,
  author       = {{ANI}},
  title        = {UP: CM Yogi Adityanath calls Lakhimpur Kheri incident 'unfortunate'},
  howpublished = {\url{https://timesofindia.indiatimes.com/india/up-cm-yogi-adityanath-calls-lakhimpur-kheri-incident-unfortunate/articleshow/86734610.cms}},
  journal      = {The Times of India},
  year         = {2021},
  month        = {October},
  day          = {3},
  note         = {Accessed: 2026-04-22}
}

@inproceedings{wang2025bias,
  title={Bias amplification: Large language models as increasingly biased media},
  author={Wang, Ze and Wu, Zekun and Zhang, Yichi and Guan, Xin and Jain, Navya and Lu, Qinyang and Gupta, Saloni and Koshiyama, Adriano},
  booktitle={Proceedings of the 14th International Joint Conference on Natural Language Processing and the 4th Conference of the Asia-Pacific Chapter of the Association for Computational Linguistics},
  pages={115--132},
  year={2025}
}

@inproceedings{sen2022analysis,
  title={Analysis of media bias in policy discourse in india},
  author={Sen, Anirban and Ghatak, Debanjan and Khanuja, Gurjeet and Rekha, Kumari and Gupta, Mehak and Dhakate, Sanket and Sharma, Kartikeya and Seth, Aaditeshwar},
  booktitle={Proceedings of the 5th ACM SIGCAS/SIGCHI Conference on Computing and Sustainable Societies},
  pages={57--77},
  year={2022}
}

@INPROCEEDINGS{9381344,
  author={Sharma, Ankur and Kaur, Navreet and Sen, Anirban and Seth, Aaditeshwar},
  booktitle={2020 IEEE/ACM International Conference on Advances in Social Networks Analysis and Mining (ASONAM)}, 
  title={Ideology Detection in the Indian Mass Media}, 
  year={2020},
  volume={},
  number={},
  pages={627-634},
  doi={10.1109/ASONAM49781.2020.9381344}}

@inproceedings{hamborg2021newsmtsc,
  title={NewsMTSC: a dataset for (multi-) target-dependent sentiment classification in political news articles},
  author={Hamborg, Felix and Donnay, Karsten},
  booktitle={Proceedings of the 16th Conference of the European Chapter of the Association for Computational Linguistics: Main Volume},
  pages={1663--1675},
  year={2021}
}

@inproceedings{traag2016structure,
  title={Structure of a media co-occurrence network},
  author={Traag, Vincent A and Reinanda, Ridho and van Klinken, Gerry},
  booktitle={Proceedings of ECCS 2014: European Conference on Complex Systems},
  pages={81--91},
  year={2016},
  organization={Springer}
}

@article{lee2025network,
  title={Network analysis reveals news press landscape and asymmetric user polarization},
  author={Lee, Byunghwee and Ryu, Hyo-sun and Lee, Jae Kook and Jeong, Hawoong and Kim, Beom Jun},
  journal={Physica A: Statistical Mechanics and its Applications},
  pages={130842},
  year={2025},
  publisher={Elsevier}
}

@article{LOW2022126722,
title = {Discerning media bias within a network of political allies and opponents: The idealized example of a biased coin},
journal = {Physica A: Statistical Mechanics and its Applications},
volume = {590},
pages = {126722},
year = {2022},
issn = {0378-4371},
doi = {https://doi.org/10.1016/j.physa.2021.126722},
url = {https://www.sciencedirect.com/science/article/pii/S037843712100933X},
author = {Nicholas Kah Yean Low and Andrew Melatos}
}

@article{https://doi.org/10.1111/soc4.12779,
author = {Segev, Elad},
title = {Textual network analysis: Detecting prevailing themes and biases in international news and social media},
journal = {Sociology Compass},
volume = {14},
number = {4},
pages = {e12779},
doi = {https://doi.org/10.1111/soc4.12779},
url = {https://compass.onlinelibrary.wiley.com/doi/abs/10.1111/soc4.12779},
eprint = {https://compass.onlinelibrary.wiley.com/doi/pdf/10.1111/soc4.12779},
year = {2020}
}

@article{wl1990toward,
  title={Toward a theory of press-state relations in the United States},
  author={WL, Bennett},
  journal={Journal of Communication},
  volume={40},
  number={2},
  pages={103--125},
  year={1990}
}

@article{flamino2023political,
  title={Political polarization of news media and influencers on Twitter in the 2016 and 2020 US presidential elections},
  author={Flamino, James and Galeazzi, Alessandro and Feldman, Stuart and Macy, Michael W and Cross, Brendan and Zhou, Zhenkun and Serafino, Matteo and Bovet, Alexandre and Makse, Hern{\'a}n A and Szymanski, Boleslaw K},
  journal={Nature Human Behaviour},
  volume={7},
  number={6},
  pages={904--916},
  year={2023},
  publisher={Nature Publishing Group UK London}
}

@article{cicchini2022news,
  title={News sharing on Twitter reveals emergent fragmentation of media agenda and persistent polarization},
  author={Cicchini, Tomas and Del Pozo, Sofia Morena and Tagliazucchi, Enzo and Balenzuela, Pablo},
  journal={EPJ Data Science},
  volume={11},
  number={1},
  pages={48},
  year={2022},
  publisher={Springer}
}

@electronic{elasticsearch,
  author = {{elasticsearch}},
  title = {elasticsearch/elasticsearch},
  url = {https://github.com/elasticsearch/elasticsearch},
  year = 2015
}

@article{honnibal2020spacy,
  author = {Honnibal, Matthew and Montani, Ines and Van Landeghem, Sofie and Boyd, Adriane},
  doi = {10.5281/zenodo.1212303},
  title = {{spaCy: Industrial-strength Natural Language Processing in Python}},
  year = 2020
}

@article{budak16,
    author = {Budak, Ceren and Goel, Sharad and Rao, Justin M.},
    title = {Fair and Balanced? Quantifying Media Bias through Crowdsourced Content Analysis},
    journal = {Public Opinion Quarterly},
    volume = {80},
    number = {S1},
    pages = {250-271},
    year = {2016},
    month = {04},
    issn = {0033-362X},
    doi = {10.1093/poq/nfw007},
    url = {https://doi.org/10.1093/poq/nfw007},
}

@article{baron06,
title = {Persistent media bias},
journal = {Journal of Public Economics},
volume = {90},
number = {1},
pages = {1-36},
year = {2006},
issn = {0047-2727},
doi = {https://doi.org/10.1016/j.jpubeco.2004.10.006},
url = {https://www.sciencedirect.com/science/article/pii/S0047272705000216},
author = {David P. Baron}
}

@misc{hamilton18,
      title={Inductive Representation Learning on Large Graphs}, 
      author={William L. Hamilton and Rex Ying and Jure Leskovec},
      year={2018},
      eprint={1706.02216},
      archivePrefix={arXiv},
      primaryClass={cs.SI},
      url={https://arxiv.org/abs/1706.02216}, 
}

@misc{grover16,
      title={node2vec: Scalable Feature Learning for Networks}, 
      author={Aditya Grover and Jure Leskovec},
      year={2016},
      eprint={1607.00653},
      archivePrefix={arXiv},
      primaryClass={cs.SI},
      url={https://arxiv.org/abs/1607.00653}, 
}

@article{CHEN2025103471,
title = {How do political connections of firms matter during an economic crisis?},
journal = {Journal of Development Economics},
volume = {175},
pages = {103471},
year = {2025},
issn = {0304-3878},
doi = {https://doi.org/10.1016/j.jdeveco.2025.103471},
url = {https://www.sciencedirect.com/science/article/pii/S0304387825000227},
author = {Yutong Chen and Gaurav Chiplunkar and Sheetal Sekhri and Anirban Sen and Aaditeshwar Seth}
}

@article{po2003news,
  title={News and its communicative quality: the inverted pyramid-when and why did it appear?},
  author={Po{\"{}} ttker, Horst},
  journal={Journalism Studies},
  volume={4},
  number={4},
  pages={501--511},
  year={2003},
  publisher={Taylor \& Francis}
}

@article{elasticsearch2018elasticsearch,
  title={Elasticsearch},
  author={Elasticsearch, BV},
  journal={software], version},
  volume={6},
  number={1},
  year={2018}
}

@article{traag2019louvain,
  title={From Louvain to Leiden: guaranteeing well-connected communities},
  author={Traag, Vincent A and Waltman, Ludo and Van Eck, Nees Jan},
  journal={Scientific reports},
  volume={9},
  number={1},
  pages={5233},
  year={2019},
  publisher={Nature Publishing Group UK London}
}

@article{benjelloun2009swoosh,
  title={Swoosh: a generic approach to entity resolution},
  author={Benjelloun, Omar and Garcia-Molina, Hector and Menestrina, David and Su, Qi and Whang, Steven Euijong and Widom, Jennifer},
  journal={The VLDB Journal},
  volume={18},
  number={1},
  pages={255--276},
  year={2009},
  publisher={Springer}
}
\section{Appendix}
\subsection{Detailed Experimental Results}
We report here our detailed experimental results for the link prediction experiments using GraphSAGE, across the four news-sources considered. We test the link prediction model under the four different experimental settings as discussed in the methodology, i.e., the supervised setting without and with structural properties included (1A and 1B), and unsupervised setting without and with structural properties included (2A and 2B). Under each setting, we also test the effect of various values of edge weight thresholds ($>2$ and $>3$) and that of incremental (rolling) and one-time training. These results are shown in tables \ref{tab:combined_toi}, \ref{tab:combined_ie}, \ref{tab:combined_dna}, and \ref{tab:combined_firstpost}.

Our findings in the main paper (tables \ref{tab:allweights} and \ref{tab:freqweights}) are a subset of the findings reported in this section. In the main paper, we observe the superior performance of link prediction for edge weight threshold $> 2$. We can see that similar findings hold even for a threshold $>3$, when compared to the \textit{no threshold} setting.

Additionally, we can see here the effect of incremental augmentation of structural features (three types of centralities and the community IDs) to the input, for the supervised approach. While the performance boost owing to the addition of these individual features is not consistent, they together help in enhancing the link prediction performance in the \textit{no threshold} setting. The incremental augmentation of structural features is not shown here for the unsupervised case, since the performance improvement was not significant, and the unsupervised approach did not outperform the supervised approach in any of the settings (no threshold and threshold applied).

Finally, while we report here the incremental (rolling window) and one-time training results, we primarily focus on the one-time results in the main paper, since incremental training and testing led to the presence of very few links in the test set (considering one month of test data for each shift of the training window). Hence, the findings around incremental training are subject to further analysis. However, consistent with the findings in the main paper, we see that the supervised approach trumps the unsupervised approach even for the incremental setting, with both accuracy and F1 score of the prediction exercise significantly higher for the supervised approach. 

\begin{table*}[h]
                 
\centering
\caption{ Experiment Results - TOI}
\label{tab:combined_toi}                 
\begin{tabular}{|cccccc|}
\hline
\multicolumn{1}{|c|}{\textbf{Training Strategy}} &
  \multicolumn{1}{c|}{\textbf{Subset Description}} &
  \multicolumn{1}{c|}{\textbf{Accuracy}} &
  \multicolumn{1}{c|}{\textbf{F1 Score}} &
  \multicolumn{1}{c|}{\textbf{Precision}} &
  \textbf{Recall} \\ \hline
\multicolumn{6}{|c|}{\textbf{Experiment 1A: Supervised (No Properties)}} \\ \hline
\multicolumn{1}{|c|}{\multirow{5}{*}{Incremental}} &
  \multicolumn{1}{c|}{August} &
  \multicolumn{1}{c|}{0.7018} &
  \multicolumn{1}{c|}{0.7547} &
  \multicolumn{1}{c|}{0.6410} &
  0.9174 \\ \cline{2-6}
\multicolumn{1}{|c|}{} &
  \multicolumn{1}{c|}{September} &
  \multicolumn{1}{c|}{0.7042} &
  \multicolumn{1}{c|}{0.7627} &
  \multicolumn{1}{c|}{0.6368} &
  0.9507 \\ \cline{2-6}
\multicolumn{1}{|c|}{} &
  \multicolumn{1}{c|}{October} &
  \multicolumn{1}{c|}{0.6870} &
  \multicolumn{1}{c|}{0.7477} &
  \multicolumn{1}{c|}{0.6263} &
  0.9275 \\ \cline{2-6}
\multicolumn{1}{|c|}{} &
  \multicolumn{1}{c|}{November} &
  \multicolumn{1}{c|}{0.6541} &
  \multicolumn{1}{c|}{0.7333} &
  \multicolumn{1}{c|}{0.5966} &
  0.9514 \\ \cline{2-6}
\multicolumn{1}{|c|}{} &
  \multicolumn{1}{c|}{December} &
  \multicolumn{1}{c|}{0.6588} &
  \multicolumn{1}{c|}{0.7171} &
  \multicolumn{1}{c|}{0.6124} &
  0.8649 \\ \hline
\multicolumn{1}{|c|}{\multirow{3}{*}{One-Time}} &
  \multicolumn{1}{c|}{All Weights} &
  \multicolumn{1}{c|}{0.6841} &
  \multicolumn{1}{c|}{0.7454} &
  \multicolumn{1}{c|}{0.6242} &
  0.9250 \\ \cline{2-6}
\multicolumn{1}{|c|}{} &
  \multicolumn{1}{c|}{Weight $>$ 2} &
  \multicolumn{1}{c|}{0.9517} &
  \multicolumn{1}{c|}{0.9497} &
  \multicolumn{1}{c|}{0.9895} &
  0.9130 \\ \cline{2-6}
\multicolumn{1}{|c|}{} &
  \multicolumn{1}{c|}{\textbf{Weight $>$ 3}} &
  \multicolumn{1}{c|}{\textbf{0.9804}} &
  \multicolumn{1}{c|}{\textbf{0.9802}} &
  \multicolumn{1}{c|}{\textbf{0.9900}} &
  \textbf{0.9706} \\ \hline
\multicolumn{6}{|c|}{\textbf{Experiment 1B: Unsupervised (No Properties)}} \\ \hline
\multicolumn{1}{|c|}{\multirow{5}{*}{Incremental}} &
  \multicolumn{1}{c|}{August} &
  \multicolumn{1}{c|}{0.6606} &
  \multicolumn{1}{c|}{0.6300} &
  \multicolumn{1}{c|}{0.6923} &
  0.5780 \\ \cline{2-6}
\multicolumn{1}{|c|}{} &
  \multicolumn{1}{c|}{September} &
  \multicolumn{1}{c|}{0.6408} &
  \multicolumn{1}{c|}{0.6946} &
  \multicolumn{1}{c|}{0.6042} &
  0.8169 \\ \cline{2-6}
\multicolumn{1}{|c|}{} &
  \multicolumn{1}{c|}{October} &
  \multicolumn{1}{c|}{0.6088} &
  \multicolumn{1}{c|}{0.6782} &
  \multicolumn{1}{c|}{0.5760} &
  0.8244 \\ \cline{2-6}
\multicolumn{1}{|c|}{} &
  \multicolumn{1}{c|}{November} &
  \multicolumn{1}{c|}{0.5919} &
  \multicolumn{1}{c|}{0.7022} &
  \multicolumn{1}{c|}{0.5528} &
  0.9622 \\ \cline{2-6}
\multicolumn{1}{|c|}{} &
  \multicolumn{1}{c|}{December} &
  \multicolumn{1}{c|}{0.7297} &
  \multicolumn{1}{c|}{0.7222} &
  \multicolumn{1}{c|}{0.7429} &
  0.7027 \\ \hline
\multicolumn{1}{|c|}{\multirow{3}{*}{One-Time}} &
  \multicolumn{1}{c|}{All Weights} &
  \multicolumn{1}{c|}{0.5509} &
  \multicolumn{1}{c|}{0.6766} &
  \multicolumn{1}{c|}{0.5286} &
  0.9398 \\ \cline{2-6}
\multicolumn{1}{|c|}{} & 
  \multicolumn{1}{c|}{\textbf{Weight $>$ 2}} & 
  \multicolumn{1}{c|}{\textbf{0.8889}} & 
  \multicolumn{1}{c|}{\textbf{0.8838}} & 
  \multicolumn{1}{c|}{\textbf{0.9259}} & 
  \textbf{0.8454} \\ \cline{2-6}
\multicolumn{1}{|c|}{} &
  \multicolumn{1}{c|}{Weight $>$ 3} &
  \multicolumn{1}{c|}{0.8529} &
  \multicolumn{1}{c|}{0.8684} &
  \multicolumn{1}{c|}{0.7857} &
  0.9706 \\ \hline
\multicolumn{6}{|c|}{\textbf{Experiment 2A: Supervised (Properties)}} \\ \hline
\multicolumn{1}{|c|}{\multirow{7}{*}{One-Time}} &
  \multicolumn{1}{c|}{All Weights (Add comm\_id)} &
  \multicolumn{1}{c|}{0.6968} &
  \multicolumn{1}{c|}{0.7561} &
  \multicolumn{1}{c|}{0.6324} &
  0.9398 \\ \cline{2-6}
\multicolumn{1}{|c|}{} &
  \multicolumn{1}{c|}{All Weights (Add degree)} &
  \multicolumn{1}{c|}{0.7082} &
  \multicolumn{1}{c|}{0.7553} &
  \multicolumn{1}{c|}{0.6502} &
  0.9009 \\ \cline{2-6}
\multicolumn{1}{|c|}{} &
  \multicolumn{1}{c|}{All Weights (Add eigen)} &
  \multicolumn{1}{c|}{0.7202} &
  \multicolumn{1}{c|}{0.7564} &
  \multicolumn{1}{c|}{0.6698} &
  0.8688 \\ \cline{2-6}
\multicolumn{1}{|c|}{} &
  \multicolumn{1}{c|}{All Weights (Add betweenness)} &
  \multicolumn{1}{c|}{0.7102} &
  \multicolumn{1}{c|}{0.7604} &
  \multicolumn{1}{c|}{0.6481} &
  0.9197 \\ \cline{2-6}
\multicolumn{1}{|c|}{} &
  \multicolumn{1}{c|}{Weight $>$ 1} &
  \multicolumn{1}{c|}{0.7162} &
  \multicolumn{1}{c|}{0.7580} &
  \multicolumn{1}{c|}{0.6607} &
  0.8889 \\ \cline{2-6}
\multicolumn{1}{|c|}{} &
  \multicolumn{1}{c|}{Weight $>$ 2} &
  \multicolumn{1}{c|}{0.9493} &
  \multicolumn{1}{c|}{0.9474} &
  \multicolumn{1}{c|}{0.9844} &
  0.9130 \\ \cline{2-6}
\multicolumn{1}{|c|}{} & 
  \multicolumn{1}{c|}{\textbf{Weight $>$ 3}} & 
  \multicolumn{1}{c|}{\textbf{0.9706}} & 
  \multicolumn{1}{c|}{\textbf{0.9706}} & 
  \multicolumn{1}{c|}{\textbf{0.9706}} & 
  \textbf{0.9706} \\ \hline
  
\multicolumn{6}{|c|}{\textbf{Experiment 2B: Unsupervised (Properties)}} \\ \hline
\multicolumn{1}{|c|}{\multirow{7}{*}{One-Time}} &
  \multicolumn{1}{c|}{All Weights (Add comm\_id)} &
  \multicolumn{1}{c|}{0.5602} &
  \multicolumn{1}{c|}{0.6694} &
  \multicolumn{1}{c|}{0.5363} &
  0.8902 \\ \cline{2-6}
\multicolumn{1}{|c|}{} &
  \multicolumn{1}{c|}{All Weights (Add degree)} &
  \multicolumn{1}{c|}{0.5964} &
  \multicolumn{1}{c|}{0.6801} &
  \multicolumn{1}{c|}{0.5633} &
  0.8581 \\ \cline{2-6}
\multicolumn{1}{|c|}{} &
  \multicolumn{1}{c|}{All Weights (Add eigen)} &
  \multicolumn{1}{c|}{0.6225} &
  \multicolumn{1}{c|}{0.6493} &
  \multicolumn{1}{c|}{0.6063} &
  0.6988 \\ \cline{2-6}
\multicolumn{1}{|c|}{} &
  \multicolumn{1}{c|}{All Weights (Add betweenness)} &
  \multicolumn{1}{c|}{0.5924} &
  \multicolumn{1}{c|}{0.6759} &
  \multicolumn{1}{c|}{0.5610} &
  0.8501 \\ \cline{2-6}
\multicolumn{1}{|c|}{} &
  \multicolumn{1}{c|}{Weight $>$ 1} &
  \multicolumn{1}{c|}{0.6285} &
  \multicolumn{1}{c|}{0.6726} &
  \multicolumn{1}{c|}{0.6013} &
  0.7631 \\ \cline{2-6}
\multicolumn{1}{|c|}{} &
  \multicolumn{1}{c|}{Weight $>$ 2} &
  \multicolumn{1}{c|}{0.8744} &
  \multicolumn{1}{c|}{0.8719} &
  \multicolumn{1}{c|}{0.8894} &
  0.8551 \\ \cline{2-6}
\multicolumn{1}{|c|}{} & 
  \multicolumn{1}{c|}{\textbf{Weight $>$ 3}} & 
  \multicolumn{1}{c|}{\textbf{0.9314}} & 
  \multicolumn{1}{c|}{\textbf{0.9300}} & 
  \multicolumn{1}{c|}{\textbf{0.9490}} & 
  \textbf{0.9118} \\\hline
\end{tabular}
\end{table*}

\begin{table*}[h]
\centering
\caption{ Experiment Results - IE}
\label{tab:combined_ie}
\begin{tabular}{|cccccc|}
\hline
\multicolumn{1}{|c|}{\textbf{Training Strategy}} &
  \multicolumn{1}{c|}{\textbf{Subset Description}} &
  \multicolumn{1}{c|}{\textbf{Accuracy}} &
  \multicolumn{1}{c|}{\textbf{F1 Score}} &
  \multicolumn{1}{c|}{\textbf{Precision}} &
  \textbf{Recall} \\ \hline
\multicolumn{6}{|c|}{\textbf{Experiment 1A: Supervised (No Properties)}} \\ \hline
\multicolumn{1}{|c|}{\multirow{5}{*}{Incremental}} &
  \multicolumn{1}{c|}{August} &
  \multicolumn{1}{c|}{0.6702} &
  \multicolumn{1}{c|}{0.7304} &
  \multicolumn{1}{c|}{0.6176} &
  0.8936 \\ \cline{2-6}
\multicolumn{1}{|c|}{} &
  \multicolumn{1}{c|}{September} &
  \multicolumn{1}{c|}{0.7619} &
  \multicolumn{1}{c|}{0.7959} &
  \multicolumn{1}{c|}{0.6964} &
  0.9286 \\ \cline{2-6}
\multicolumn{1}{|c|}{} &
  \multicolumn{1}{c|}{October} &
  \multicolumn{1}{c|}{0.7176} &
  \multicolumn{1}{c|}{0.7600} &
  \multicolumn{1}{c|}{0.6609} &
  0.8941 \\ \cline{2-6}
\multicolumn{1}{|c|}{} &
  \multicolumn{1}{c|}{November} &
  \multicolumn{1}{c|}{0.6905} &
  \multicolumn{1}{c|}{0.7593} &
  \multicolumn{1}{c|}{0.6212} &
  0.9762 \\ \cline{2-6}
\multicolumn{1}{|c|}{} &
  \multicolumn{1}{c|}{December} &
  \multicolumn{1}{c|}{0.7396} &
  \multicolumn{1}{c|}{0.7826} &
  \multicolumn{1}{c|}{0.6716} &
  0.9375 \\ \hline
\multicolumn{1}{|c|}{\multirow{3}{*}{One-Time}} &
  \multicolumn{1}{c|}{All Weights} &
  \multicolumn{1}{c|}{0.6409} &
  \multicolumn{1}{c|}{0.6376} &
  \multicolumn{1}{c|}{0.6435} &
  0.6318 \\ \cline{2-6}
\multicolumn{1}{|c|}{} &
  \multicolumn{1}{c|}{Weight $>$ 2} &
  \multicolumn{1}{c|}{0.9182} &
  \multicolumn{1}{c|}{0.9109} &
  \multicolumn{1}{c|}{1.0000} &
  0.8364 \\ \cline{2-6}
\multicolumn{1}{|c|}{} & 
  \multicolumn{1}{c|}{\textbf{Weight $>$ 3}} & 
  \multicolumn{1}{c|}{\textbf{0.9333}} & 
  \multicolumn{1}{c|}{\textbf{0.9286}} & 
  \multicolumn{1}{c|}{\textbf{1.0000}} & 
  \textbf{0.8667} \\ \hline
\multicolumn{6}{|c|}{\textbf{Experiment 1B: Unsupervised (No Properties)}} \\ \hline
\multicolumn{1}{|c|}{\multirow{5}{*}{Incremental}} &
  \multicolumn{1}{c|}{August} &
  \multicolumn{1}{c|}{0.6596} &
  \multicolumn{1}{c|}{0.7241} &
  \multicolumn{1}{c|}{0.6087} &
  0.8936 \\ \cline{2-6}
\multicolumn{1}{|c|}{} &
  \multicolumn{1}{c|}{September} &
  \multicolumn{1}{c|}{0.6429} &
  \multicolumn{1}{c|}{0.5714} &
  \multicolumn{1}{c|}{0.7143} &
  0.4762 \\ \cline{2-6}
\multicolumn{1}{|c|}{} &
  \multicolumn{1}{c|}{October} &
  \multicolumn{1}{c|}{0.5706} &
  \multicolumn{1}{c|}{0.5576} &
  \multicolumn{1}{c|}{0.5750} &
  0.5412 \\ \cline{2-6}
\multicolumn{1}{|c|}{} &
  \multicolumn{1}{c|}{November} &
  \multicolumn{1}{c|}{0.5357} &
  \multicolumn{1}{c|}{0.6723} &
  \multicolumn{1}{c|}{0.5195} &
  0.9524 \\ \cline{2-6}
\multicolumn{1}{|c|}{} &
  \multicolumn{1}{c|}{December} &
  \multicolumn{1}{c|}{0.7396} &
  \multicolumn{1}{c|}{0.7475} &
  \multicolumn{1}{c|}{0.7255} &
  0.7708 \\ \hline
\multicolumn{1}{|c|}{\multirow{3}{*}{One-Time}} &
  \multicolumn{1}{c|}{All Weights} &
  \multicolumn{1}{c|}{0.6295} &
  \multicolumn{1}{c|}{0.6253} &
  \multicolumn{1}{c|}{0.6326} &
  0.6182 \\ \cline{2-6}
\multicolumn{1}{|c|}{} &
  \multicolumn{1}{c|}{Weight $>$ 2} &
  \multicolumn{1}{c|}{0.8909} &
  \multicolumn{1}{c|}{0.8889} &
  \multicolumn{1}{c|}{0.9057} &
  0.8727 \\ \cline{2-6}
\multicolumn{1}{|c|}{} & 
  \multicolumn{1}{c|}{\textbf{Weight $>$ 3}} & 
  \multicolumn{1}{c|}{\textbf{0.9667}} & 
  \multicolumn{1}{c|}{\textbf{0.9655}} & 
  \multicolumn{1}{c|}{\textbf{1.0000}} & 
  \textbf{0.9333} \\ \hline
\multicolumn{6}{|c|}{\textbf{Experiment 2A: Supervised (Properties)}} \\ \hline
\multicolumn{1}{|c|}{\multirow{7}{*}{One-Time}} &
  \multicolumn{1}{c|}{All Weights (Add comm\_id)} &
  \multicolumn{1}{c|}{0.6500} &
  \multicolumn{1}{c|}{0.6468} &
  \multicolumn{1}{c|}{0.6528} &
  0.6409 \\ \cline{2-6}
\multicolumn{1}{|c|}{} &
  \multicolumn{1}{c|}{All Weights (Add degree)} &
  \multicolumn{1}{c|}{0.6659} &
  \multicolumn{1}{c|}{0.5788} &
  \multicolumn{1}{c|}{0.7829} &
  0.4591 \\ \cline{2-6}
\multicolumn{1}{|c|}{} &
  \multicolumn{1}{c|}{All Weights (Add eigen)} &
  \multicolumn{1}{c|}{0.6455} &
  \multicolumn{1}{c|}{0.5593} &
  \multicolumn{1}{c|}{0.7388} &
  0.4500 \\ \cline{2-6}
\multicolumn{1}{|c|}{} &
  \multicolumn{1}{c|}{All Weights (Add betweenness)} &
  \multicolumn{1}{c|}{0.7182} &
  \multicolumn{1}{c|}{0.7182} &
  \multicolumn{1}{c|}{0.7182} &
  0.7182 \\ \cline{2-6}
\multicolumn{1}{|c|}{} &
  \multicolumn{1}{c|}{Weight $>$ 1} &
  \multicolumn{1}{c|}{0.6568} &
  \multicolumn{1}{c|}{0.5863} &
  \multicolumn{1}{c|}{0.7379} &
  0.4864 \\ \cline{2-6}
\multicolumn{1}{|c|}{} & 
  \multicolumn{1}{c|}{\textbf{Weight $>$ 2}} & 
  \multicolumn{1}{c|}{\textbf{0.8636}} & 
  \multicolumn{1}{c|}{\textbf{0.8421}} & 
  \multicolumn{1}{c|}{\textbf{1.0000}} & 
  \textbf{0.7273} \\\cline{2-6}
\multicolumn{1}{|c|}{} &
  \multicolumn{1}{c|}{Weight $>$ 3} &
  \multicolumn{1}{c|}{0.8167} &
  \multicolumn{1}{c|}{0.7755} &
  \multicolumn{1}{c|}{1.0000} &
  0.6333 \\ \hline
\multicolumn{6}{|c|}{\textbf{Experiment 2B: Unsupervised (Properties)}} \\ \hline
\multicolumn{1}{|c|}{\multirow{7}{*}{One-Time}} &
  \multicolumn{1}{c|}{All Weights (Add comm\_id)} &
  \multicolumn{1}{c|}{0.5318} &
  \multicolumn{1}{c|}{0.6069} &
  \multicolumn{1}{c|}{0.5230} &
  0.7227 \\ \cline{2-6}
\multicolumn{1}{|c|}{} &
  \multicolumn{1}{c|}{All Weights (Add degree)} &
  \multicolumn{1}{c|}{0.6273} &
  \multicolumn{1}{c|}{0.5920} &
  \multicolumn{1}{c|}{0.6538} &
  0.5409 \\ \cline{2-6}
\multicolumn{1}{|c|}{} &
  \multicolumn{1}{c|}{All Weights (Add eigen)} &
  \multicolumn{1}{c|}{0.6659} &
  \multicolumn{1}{c|}{0.6316} &
  \multicolumn{1}{c|}{0.7039} &
  0.5727 \\ \cline{2-6}
\multicolumn{1}{|c|}{} &
  \multicolumn{1}{c|}{All Weights (Add betweenness)} &
  \multicolumn{1}{c|}{0.6364} &
  \multicolumn{1}{c|}{0.6887} &
  \multicolumn{1}{c|}{0.6020} &
  0.8045 \\ \cline{2-6}
\multicolumn{1}{|c|}{} &
  \multicolumn{1}{c|}{Weight $>$ 1} &
  \multicolumn{1}{c|}{0.6023} &
  \multicolumn{1}{c|}{0.6331} &
  \multicolumn{1}{c|}{0.5875} &
  0.6864 \\ \cline{2-6}
\multicolumn{1}{|c|}{} &
  \multicolumn{1}{c|}{Weight $>$ 2} &
  \multicolumn{1}{c|}{0.8000} &
  \multicolumn{1}{c|}{0.7708} &
  \multicolumn{1}{c|}{0.9024} &
  0.6727 \\ \cline{2-6}
\multicolumn{1}{|c|}{} & 
  \multicolumn{1}{c|}{\textbf{Weight $>$ 3}} & 
  \multicolumn{1}{c|}{\textbf{0.9833}} & 
  \multicolumn{1}{c|}{\textbf{0.9831}} & 
  \multicolumn{1}{c|}{\textbf{1.0000}} & 
  \textbf{0.9667} \\\hline

\end{tabular}
\end{table*}
\begin{table*}[h]
\centering
\caption{Experiment Results - dna}
\label{tab:combined_dna}
\begin{tabular}{|cccccc|}
\hline
\multicolumn{1}{|c|}{\textbf{Training Strategy}} &
  \multicolumn{1}{c|}{\textbf{Subset Description}} &
  \multicolumn{1}{c|}{\textbf{Accuracy}} &
  \multicolumn{1}{c|}{\textbf{F1 Score}} &
  \multicolumn{1}{c|}{\textbf{Precision}} &
  \textbf{Recall} \\ \hline
\multicolumn{6}{|c|}{\textbf{Experiment 1A: Supervised (No Properties)}} \\ \hline
\multicolumn{1}{|c|}{\multirow{5}{*}{Incremental}} &
  \multicolumn{1}{c|}{August} &
  \multicolumn{1}{c|}{0.5000} &
  \multicolumn{1}{c|}{0.0000} &
  \multicolumn{1}{c|}{0.0000} &
  0.0000 \\ \cline{2-6}
\multicolumn{1}{|c|}{} &
  \multicolumn{1}{c|}{September} &
  \multicolumn{1}{c|}{0.8750} &
  \multicolumn{1}{c|}{0.8889} &
  \multicolumn{1}{c|}{0.8000} &
  1.0000 \\ \cline{2-6}
\multicolumn{1}{|c|}{} &
  \multicolumn{1}{c|}{October} &
  \multicolumn{1}{c|}{0.7278} &
  \multicolumn{1}{c|}{0.7435} &
  \multicolumn{1}{c|}{0.7030} &
  0.7889 \\ \cline{2-6}
\multicolumn{1}{|c|}{} &
  \multicolumn{1}{c|}{November} &
  \multicolumn{1}{c|}{0.6765} &
  \multicolumn{1}{c|}{0.7442} &
  \multicolumn{1}{c|}{0.6154} &
  0.9412 \\ \cline{2-6}
\multicolumn{1}{|c|}{} &
  \multicolumn{1}{c|}{December} &
  \multicolumn{1}{c|}{0.8333} &
  \multicolumn{1}{c|}{0.8000} &
  \multicolumn{1}{c|}{1.0000} &
  0.6667 \\ \hline
\multicolumn{1}{|c|}{\multirow{3}{*}{One-Time}} &
  \multicolumn{1}{c|}{All Weights} &
  \multicolumn{1}{c|}{0.6568} &
  \multicolumn{1}{c|}{0.5091} &
  \multicolumn{1}{c|}{0.8936} &
  0.3559 \\ \cline{2-6}
\multicolumn{1}{|c|}{} & 
  \multicolumn{1}{c|}{\textbf{Weight $>$ 2}} & 
  \multicolumn{1}{c|}{\textbf{0.9394}} & 
  \multicolumn{1}{c|}{\textbf{0.9355}} & 
  \multicolumn{1}{c|}{\textbf{1.0000}} & 
  \textbf{0.8788} \\\cline{2-6}
\multicolumn{1}{|c|}{} &
  \multicolumn{1}{c|}{Weight $>$ 3} &
  \multicolumn{1}{c|}{0.8846} &
  \multicolumn{1}{c|}{0.8696} &
  \multicolumn{1}{c|}{1.0000} &
  0.7692 \\ \hline
\multicolumn{6}{|c|}{\textbf{Experiment 1B: Unsupervised (No Properties)}} \\ \hline
\multicolumn{1}{|c|}{\multirow{5}{*}{Incremental}} &
  \multicolumn{1}{c|}{August} &
  \multicolumn{1}{c|}{0.5000} &
  \multicolumn{1}{c|}{0.0000} &
  \multicolumn{1}{c|}{0.0000} &
  0.0000 \\ \cline{2-6}
\multicolumn{1}{|c|}{} &
  \multicolumn{1}{c|}{September} &
  \multicolumn{1}{c|}{0.6250} &
  \multicolumn{1}{c|}{0.5714} &
  \multicolumn{1}{c|}{0.6667} &
  0.5000 \\ \cline{2-6}
\multicolumn{1}{|c|}{} &
  \multicolumn{1}{c|}{October} &
  \multicolumn{1}{c|}{0.5278} &
  \multicolumn{1}{c|}{0.6693} &
  \multicolumn{1}{c|}{0.5150} &
  0.9556 \\ \cline{2-6}
\multicolumn{1}{|c|}{} &
  \multicolumn{1}{c|}{November} &
  \multicolumn{1}{c|}{0.6471} &
  \multicolumn{1}{c|}{0.7273} &
  \multicolumn{1}{c|}{0.5926} &
  0.9412 \\ \cline{2-6}
\multicolumn{1}{|c|}{} &
  \multicolumn{1}{c|}{December} &
  \multicolumn{1}{c|}{0.6250} &
  \multicolumn{1}{c|}{0.7097} &
  \multicolumn{1}{c|}{0.5789} &
  0.9167 \\ \hline
\multicolumn{1}{|c|}{\multirow{3}{*}{One-Time}} &
  \multicolumn{1}{c|}{All Weights} &
  \multicolumn{1}{c|}{0.6568} &
  \multicolumn{1}{c|}{0.5888} &
  \multicolumn{1}{c|}{0.7342} &
  0.4915 \\ \cline{2-6}
\multicolumn{1}{|c|}{} & 
  \multicolumn{1}{c|}{\textbf{Weight $>$ 2}} & 
  \multicolumn{1}{c|}{\textbf{0.8333}} & 
  \multicolumn{1}{c|}{\textbf{0.8070}} & 
  \multicolumn{1}{c|}{\textbf{0.9583}} & 
  \textbf{0.6970} \\\cline{2-6}
\multicolumn{1}{|c|}{} &
  \multicolumn{1}{c|}{Weight $>$ 3} &
  \multicolumn{1}{c|}{0.8077} &
  \multicolumn{1}{c|}{0.7619} &
  \multicolumn{1}{c|}{1.0000} &
  0.6154 \\ \hline
\multicolumn{6}{|c|}{\textbf{Experiment 2A: Supervised (Properties)}} \\ \hline
\multicolumn{1}{|c|}{\multirow{7}{*}{One-Time}} &
  \multicolumn{1}{c|}{All Weights (Add comm\_id)} &
  \multicolumn{1}{c|}{0.6907} &
  \multicolumn{1}{c|}{0.6332} &
  \multicolumn{1}{c|}{0.7778} &
  0.5339 \\ \cline{2-6}
\multicolumn{1}{|c|}{} &
  \multicolumn{1}{c|}{All Weights (Add degree)} &
  \multicolumn{1}{c|}{0.6822} &
  \multicolumn{1}{c|}{0.5455} &
  \multicolumn{1}{c|}{0.9574} &
  0.3814 \\ \cline{2-6}
\multicolumn{1}{|c|}{} &
  \multicolumn{1}{c|}{All Weights (Add eigen)} &
  \multicolumn{1}{c|}{0.6653} &
  \multicolumn{1}{c|}{0.5031} &
  \multicolumn{1}{c|}{0.9756} &
  0.3390 \\ \cline{2-6}
\multicolumn{1}{|c|}{} &
  \multicolumn{1}{c|}{All Weights (Add betweenness)} &
  \multicolumn{1}{c|}{0.6695} &
  \multicolumn{1}{c|}{0.5185} &
  \multicolumn{1}{c|}{0.9545} &
  0.3559 \\ \cline{2-6}
\multicolumn{1}{|c|}{} &
  \multicolumn{1}{c|}{Weight $>$ 1} &
  \multicolumn{1}{c|}{0.6695} &
  \multicolumn{1}{c|}{0.5063} &
  \multicolumn{1}{c|}{1.0000} &
  0.3390 \\ \cline{2-6}
\multicolumn{1}{|c|}{} &
  \multicolumn{1}{c|}{Weight $>$ 2} &
  \multicolumn{1}{c|}{0.8788} &
  \multicolumn{1}{c|}{0.8621} &
  \multicolumn{1}{c|}{1.0000} &
  0.7576 \\ \cline{2-6}
\multicolumn{1}{|c|}{} & 
  \multicolumn{1}{c|}{\textbf{Weight $>$ 3}} & 
  \multicolumn{1}{c|}{\textbf{0.9615}} & 
  \multicolumn{1}{c|}{\textbf{0.9600}} & 
  \multicolumn{1}{c|}{\textbf{1.0000}} & 
  \textbf{0.9231} \\ \hline
  
\multicolumn{6}{|c|}{\textbf{Experiment 2B: Unsupervised (Properties)}} \\ \hline
\multicolumn{1}{|c|}{\multirow{7}{*}{One-Time}} &
  \multicolumn{1}{c|}{All Weights (Add comm\_id)} &
  \multicolumn{1}{c|}{0.6525} &
  \multicolumn{1}{c|}{0.5119} &
  \multicolumn{1}{c|}{0.8600} &
  0.3644 \\ \cline{2-6}
\multicolumn{1}{|c|}{} &
  \multicolumn{1}{c|}{All Weights (Add degree)} &
  \multicolumn{1}{c|}{0.6864} &
  \multicolumn{1}{c|}{0.5647} &
  \multicolumn{1}{c|}{0.9231} &
  0.4068 \\ \cline{2-6}
\multicolumn{1}{|c|}{} &
  \multicolumn{1}{c|}{All Weights (Add eigen)} &
  \multicolumn{1}{c|}{0.6949} &
  \multicolumn{1}{c|}{0.5814} &
  \multicolumn{1}{c|}{0.9259} &
  0.4237 \\ \cline{2-6}
\multicolumn{1}{|c|}{} &
  \multicolumn{1}{c|}{All Weights (Add betweenness)} &
  \multicolumn{1}{c|}{0.6271} &
  \multicolumn{1}{c|}{0.4286} &
  \multicolumn{1}{c|}{0.9167} &
  0.2797 \\ \cline{2-6}
\multicolumn{1}{|c|}{} &
  \multicolumn{1}{c|}{Weight $>$ 1} &
  \multicolumn{1}{c|}{0.6864} &
  \multicolumn{1}{c|}{0.5889} &
  \multicolumn{1}{c|}{0.8548} &
  0.4492 \\ \cline{2-6}
\multicolumn{1}{|c|}{} &
  \multicolumn{1}{c|}{Weight $>$ 2} &
  \multicolumn{1}{c|}{0.8333} &
  \multicolumn{1}{c|}{0.8197} &
  \multicolumn{1}{c|}{0.8929} &
  0.7576 \\ \cline{2-6}
\multicolumn{1}{|c|}{} & 
  \multicolumn{1}{c|}{\textbf{Weight $>$ 3}} & 
  \multicolumn{1}{c|}{\textbf{0.9231}} & 
  \multicolumn{1}{c|}{\textbf{0.9231}} & 
  \multicolumn{1}{c|}{\textbf{0.9231}} & 
  \textbf{0.9231} \\ \hline

\end{tabular}
\end{table*}

\begin{table*}[h]
\centering
\caption{ Experiment Results - firstpost}
\label{tab:combined_firstpost}
\begin{tabular}{|cccccc|}
\hline
\multicolumn{1}{|c|}{\textbf{Training Strategy}} &
  \multicolumn{1}{c|}{\textbf{Subset Description}} &
  \multicolumn{1}{c|}{\textbf{Accuracy}} &
  \multicolumn{1}{c|}{\textbf{F1 Score}} &
  \multicolumn{1}{c|}{\textbf{Precision}} &
  \textbf{Recall} \\ \hline
\multicolumn{6}{|c|}{\textbf{Experiment 1A: Supervised (No Properties)}} \\ \hline
\multicolumn{1}{|c|}{\multirow{5}{*}{Incremental}} &
  \multicolumn{1}{c|}{August} &
  \multicolumn{1}{c|}{0.8045} &
  \multicolumn{1}{c|}{0.8182} &
  \multicolumn{1}{c|}{0.7647} &
  0.8797 \\ \cline{2-6}
\multicolumn{1}{|c|}{} &
  \multicolumn{1}{c|}{September} &
  \multicolumn{1}{c|}{0.6705} &
  \multicolumn{1}{c|}{0.7339} &
  \multicolumn{1}{c|}{0.6154} &
  0.9091 \\ \cline{2-6}
\multicolumn{1}{|c|}{} &
  \multicolumn{1}{c|}{October} &
  \multicolumn{1}{c|}{0.7465} &
  \multicolumn{1}{c|}{0.6897} &
  \multicolumn{1}{c|}{0.8889} &
  0.5634 \\ \cline{2-6}
\multicolumn{1}{|c|}{} &
  \multicolumn{1}{c|}{November} &
  \multicolumn{1}{c|}{0.6639} &
  \multicolumn{1}{c|}{0.7324} &
  \multicolumn{1}{c|}{0.6084} &
  0.9197 \\ \cline{2-6}
\multicolumn{1}{|c|}{} &
  \multicolumn{1}{c|}{December} &
  \multicolumn{1}{c|}{0.7238} &
  \multicolumn{1}{c|}{0.7711} &
  \multicolumn{1}{c|}{0.6584} &
  0.9302 \\ \hline
\multicolumn{1}{|c|}{\multirow{3}{*}{One-Time}} &
  \multicolumn{1}{c|}{All Weights} &
  \multicolumn{1}{c|}{0.7917} &
  \multicolumn{1}{c|}{0.7766} &
  \multicolumn{1}{c|}{0.8371} &
  0.7242 \\ \cline{2-6}
\multicolumn{1}{|c|}{} &
  \multicolumn{1}{c|}{Weight $>$ 2} &
  \multicolumn{1}{c|}{0.9648} &
  \multicolumn{1}{c|}{0.9647} &
  \multicolumn{1}{c|}{0.9685} &
  0.9609 \\ \cline{2-6}
\multicolumn{1}{|c|}{} & 
  \multicolumn{1}{c|}{\textbf{Weight $>$ 3}} & 
  \multicolumn{1}{c|}{\textbf{0.9895}} & 
  \multicolumn{1}{c|}{\textbf{0.9894}} & 
  \multicolumn{1}{c|}{\textbf{1.0000}} & 
  \textbf{0.9790} \\ \hline
\multicolumn{6}{|c|}{\textbf{Experiment 1B: Unsupervised (No Properties)}} \\ \hline
\multicolumn{1}{|c|}{\multirow{5}{*}{Incremental}} &
  \multicolumn{1}{c|}{August} &
  \multicolumn{1}{c|}{0.7707} &
  \multicolumn{1}{c|}{0.7875} &
  \multicolumn{1}{c|}{0.7338} &
  0.8496 \\ \cline{2-6}
\multicolumn{1}{|c|}{} &
  \multicolumn{1}{c|}{September} &
  \multicolumn{1}{c|}{0.7614} &
  \multicolumn{1}{c|}{0.7789} &
  \multicolumn{1}{c|}{0.7255} &
  0.8409 \\ \cline{2-6}
\multicolumn{1}{|c|}{} &
  \multicolumn{1}{c|}{October} &
  \multicolumn{1}{c|}{0.6620} &
  \multicolumn{1}{c|}{0.6923} &
  \multicolumn{1}{c|}{0.6353} &
  0.7606 \\ \cline{2-6}
\multicolumn{1}{|c|}{} &
  \multicolumn{1}{c|}{November} &
  \multicolumn{1}{c|}{0.5970} &
  \multicolumn{1}{c|}{0.7128} &
  \multicolumn{1}{c|}{0.5537} &
  1.0000 \\ \cline{2-6}
\multicolumn{1}{|c|}{} &
  \multicolumn{1}{c|}{December} &
  \multicolumn{1}{c|}{0.6599} &
  \multicolumn{1}{c|}{0.7310} &
  \multicolumn{1}{c|}{0.6046} &
  0.9244 \\ \hline
\multicolumn{1}{|c|}{\multirow{3}{*}{One-Time}} &
  \multicolumn{1}{c|}{All Weights} &
  \multicolumn{1}{c|}{0.7356} &
  \multicolumn{1}{c|}{0.7250} &
  \multicolumn{1}{c|}{0.7553} &
  0.6970 \\ \cline{2-6}
\multicolumn{1}{|c|}{} & 
  \multicolumn{1}{c|}{\textbf{Weight $>$ 2}} & 
  \multicolumn{1}{c|}{\textbf{0.9395}} & 
  \multicolumn{1}{c|}{\textbf{0.9412}} & 
  \multicolumn{1}{c|}{\textbf{0.9151}} & 
  \textbf{0.9688} \\ \cline{2-6}
\multicolumn{1}{|c|}{} &
  \multicolumn{1}{c|}{Weight $>$ 3} &
  \multicolumn{1}{c|}{0.9266} &
  \multicolumn{1}{c|}{0.9231} &
  \multicolumn{1}{c|}{0.9692} &
  0.8811 \\ \hline
\multicolumn{6}{|c|}{\textbf{Experiment 2A: Supervised (Properties)}} \\ \hline
\multicolumn{1}{|c|}{\multirow{7}{*}{One-Time}} &
  \multicolumn{1}{c|}{All Weights (Add comm\_id)} &
  \multicolumn{1}{c|}{0.7811} &
  \multicolumn{1}{c|}{0.7652} &
  \multicolumn{1}{c|}{0.8249} &
  0.7136 \\ \cline{2-6}
\multicolumn{1}{|c|}{} &
  \multicolumn{1}{c|}{All Weights (Add degree)} &
  \multicolumn{1}{c|}{0.8008} &
  \multicolumn{1}{c|}{0.7972} &
  \multicolumn{1}{c|}{0.8116} &
  0.7833 \\ \cline{2-6}
\multicolumn{1}{|c|}{} &
  \multicolumn{1}{c|}{All Weights (Add eigen)} &
  \multicolumn{1}{c|}{0.7924} &
  \multicolumn{1}{c|}{0.7964} &
  \multicolumn{1}{c|}{0.7813} &
  0.8121 \\ \cline{2-6}
\multicolumn{1}{|c|}{} &
  \multicolumn{1}{c|}{All Weights (Add betweenness)} &
  \multicolumn{1}{c|}{0.7803} &
  \multicolumn{1}{c|}{0.7823} &
  \multicolumn{1}{c|}{0.7753} &
  0.7894 \\ \cline{2-6}
\multicolumn{1}{|c|}{} &
  \multicolumn{1}{c|}{Weight $>$ 1} &
  \multicolumn{1}{c|}{0.7947} &
  \multicolumn{1}{c|}{0.7967} &
  \multicolumn{1}{c|}{0.7890} &
  0.8045 \\ \cline{2-6}
\multicolumn{1}{|c|}{} &
  \multicolumn{1}{c|}{Weight $>$ 2} &
  \multicolumn{1}{c|}{0.9648} &
  \multicolumn{1}{c|}{0.9641} &
  \multicolumn{1}{c|}{0.9837} &
  0.9453 \\ \cline{2-6}
\multicolumn{1}{|c|}{} & 
  \multicolumn{1}{c|}{\textbf{Weight $>$ 3}} & 
  \multicolumn{1}{c|}{\textbf{0.9895}} & 
  \multicolumn{1}{c|}{\textbf{0.9894}} & 
  \multicolumn{1}{c|}{\textbf{1.0000}} & 
  \textbf{0.9790} \\  \hline
  
\multicolumn{6}{|c|}{\textbf{Experiment 2B: Unsupervised (Properties)}} \\ \hline
\multicolumn{1}{|c|}{\multirow{7}{*}{One-Time}} &
  \multicolumn{1}{c|}{All Weights (Add comm\_id)} &
  \multicolumn{1}{c|}{0.7545} &
  \multicolumn{1}{c|}{0.7481} &
  \multicolumn{1}{c|}{0.7684} &
  0.7288 \\ \cline{2-6}
\multicolumn{1}{|c|}{} &
  \multicolumn{1}{c|}{All Weights (Add degree)} &
  \multicolumn{1}{c|}{0.7720} &
  \multicolumn{1}{c|}{0.7353} &
  \multicolumn{1}{c|}{0.8763} &
  0.6333 \\ \cline{2-6}
\multicolumn{1}{|c|}{} &
  \multicolumn{1}{c|}{All Weights (Add eigen)} &
  \multicolumn{1}{c|}{0.7545} &
  \multicolumn{1}{c|}{0.7578} &
  \multicolumn{1}{c|}{0.7478} &
  0.7682 \\ \cline{2-6}
\multicolumn{1}{|c|}{} &
  \multicolumn{1}{c|}{All Weights (Add betweenness)} &
  \multicolumn{1}{c|}{0.7621} &
  \multicolumn{1}{c|}{0.7543} &
  \multicolumn{1}{c|}{0.7799} &
  0.7303 \\ \cline{2-6}
\multicolumn{1}{|c|}{} &
  \multicolumn{1}{c|}{Weight $>$ 1} &
  \multicolumn{1}{c|}{0.7871} &
  \multicolumn{1}{c|}{0.7743} &
  \multicolumn{1}{c|}{0.8239} &
  0.7303 \\ \cline{2-6}
\multicolumn{1}{|c|}{} &
  \multicolumn{1}{c|}{Weight $>$ 2} &
  \multicolumn{1}{c|}{0.9004} &
  \multicolumn{1}{c|}{0.8994} &
  \multicolumn{1}{c|}{0.9084} &
  0.8906 \\ \cline{2-6}
\multicolumn{1}{|c|}{} & 
  \multicolumn{1}{c|}{\textbf{Weight $>$ 3}} & 
  \multicolumn{1}{c|}{\textbf{0.9545}} & 
  \multicolumn{1}{c|}{\textbf{0.9541}} & 
  \multicolumn{1}{c|}{\textbf{0.9643}} & 
  \textbf{0.9441} \\ \hline
\end{tabular}
\end{table*}

\subsection{Comparison with Baselines}
Co-occurrence link predictability is proposed as one of the axes of analysis of reporting preferences in news media, in this study. Our findings suggest that for all of the four news-sources, the link predictability is non-trivial. The predictability also varies across sources, indicative of variation in reporting preferences.

We further assess the robustness of these findings by comparing the GraphSAGE-based link predictor with standard baseline models. These models are:
\begin{itemize}
    \item \textbf{Random Baseline}: We construct a random baseline by independently assigning the presence or absence of an edge between each pair of entities in the test set with uniform probability, and predict a link when the sampled probability exceeds 0.5. This experiment is iterated 100 times, and the mean accuracy and F1 score are reported in table \ref{tab:random_baseline}.
    \item \textbf{Community ID-based Prediction}: In this case, we predict a link between two entities in the test data if they have the same community ID (Leiden). The underlying assumption here is that two nodes in the same community have a greater probability of co-occur than those in different communities. The accuracy and F1 scores are reported in table \ref{tab:communityidbaseline}
\end{itemize}
We can see that for both approaches, the F1 score is substantially lower for the baselines compared to the GraphSAGE-based link prediction model. This indicates that the observed performance is not driven by trivial (community-based) or random patterns, but rather by meaningful structural signals captured by the model.

\begin{table*}[h]
\centering

\begin{minipage}{0.48\textwidth}
\centering
\caption{Random Baseline Results}
\label{tab:random_baseline}
\begin{tabular}{|ccccc|}
\hline
\multicolumn{1}{|c|}{\textbf{Media House}} &
  \multicolumn{1}{c|}{\textbf{Accuracy}} &
  \multicolumn{1}{c|}{\textbf{F1 Score}} &
  \multicolumn{1}{c|}{\textbf{Precision}} &
  \textbf{Recall} \\ \hline

\multicolumn{1}{|c|}{dna}       & \multicolumn{1}{c|}{0.5001} & \multicolumn{1}{c|}{0.0250} & \multicolumn{1}{c|}{0.0128} & 0.5005 \\ \hline
\multicolumn{1}{|c|}{firstpost} & \multicolumn{1}{c|}{0.5000} & \multicolumn{1}{c|}{0.0283} & \multicolumn{1}{c|}{0.0146} & 0.5002 \\ \hline
\multicolumn{1}{|c|}{IE}        & \multicolumn{1}{c|}{0.5000} & \multicolumn{1}{c|}{0.0035} & \multicolumn{1}{c|}{0.0018} & 0.5008 \\ \hline
\multicolumn{1}{|c|}{TOI}       & \multicolumn{1}{c|}{0.5000} & \multicolumn{1}{c|}{0.0010} & \multicolumn{1}{c|}{0.0005} & 0.5002 \\ \hline
\end{tabular}
\end{minipage}
\hfill
\begin{minipage}{0.48\textwidth}
\centering
\caption{Community ID baseline Results}
\label{tab:communityidbaseline}
\begin{tabular}{|ccccc|}
\hline
\multicolumn{1}{|c|}{\textbf{Media House}} &
  \multicolumn{1}{c|}{\textbf{Accuracy}} &
  \multicolumn{1}{c|}{\textbf{F1 Score}} &
  \multicolumn{1}{c|}{\textbf{Precision}} &
  \textbf{Recall} \\ \hline

\multicolumn{1}{|c|}{dna}       & \multicolumn{1}{c|}{0.9051} & \multicolumn{1}{c|}{0.1720} & \multicolumn{1}{c|}{0.0968} & 0.7682 \\ \hline
\multicolumn{1}{|c|}{firstpost} & \multicolumn{1}{c|}{0.8561} & \multicolumn{1}{c|}{0.1524} & \multicolumn{1}{c|}{0.0834} & 0.8877 \\ \hline
\multicolumn{1}{|c|}{IE}        & \multicolumn{1}{c|}{0.9734} & \multicolumn{1}{c|}{0.1008} & \multicolumn{1}{c|}{0.0536} & 0.8398 \\ \hline
\multicolumn{1}{|c|}{TOI}       & \multicolumn{1}{c|}{0.9875} & \multicolumn{1}{c|}{0.0664} & \multicolumn{1}{c|}{0.0345} & 0.8657 \\ \hline

\end{tabular}
\end{minipage}

\end{table*}

\end{document}